\documentclass[useAMS,usenatbib]{mn2e}
\usepackage{listings}
\usepackage{graphics}
\usepackage{subfigure}
\usepackage{graphicx}
\usepackage{amssymb}
\usepackage{hyperref}
\usepackage{aas_macros}

\title[
Host galaxy$-$active nucleus alignments in the SDSS
]{Host galaxy$-$active galactic nucleus alignments in the SDSS-DR7} 
\author[Lagos et al.]{
\parbox[t]{\textwidth}{
\vspace{-1.0cm}
Claudia del P. Lagos$^{1,2}$,
Nelson D. Padilla$^{2,3}$,
Michael A. Strauss$^{4}$,
Sof\'ia A. Cora$^{5,6}$,
Lei Hao$^{7,8}$
}
\vspace*{6pt} \\
$^{1}$Institute for Computational Cosmology, Department of Physics,
University of Durham, South Road, Durham, DH1 3LE, UK. \\
$^{2}$Departamento Astronom\'ia y Astrof\'isica, Pontificia Universidad
Cat\'olica de Chile, Av. Vicu\~na Mackenna 4860, Stgo., Chile\\
$^{3}$Centro de Astro-Ingenier\'\i a, Pontificia Universidad
Cat\'olica de Chile, Av. Vicu\~na Mackenna 4860, Stgo., Chile\\
$^{4}$Department of Astrophysical Sciences, Princeton University,
Princeton, NJ 08544, USA\\
$^{5}$Facultad de Ciencias Astron\'omicas y Geof\'isicas de la Universidad
Nacional de La Plata, and Instituto de Astrof\'isica de La Plata \\ (CCT La Plata,
CONICET, UNLP), Observatorio Astron\'omico, Paseo del Bosque S/N, 1900 La Plata,
Argentina\\
$^{6}$Consejo Nacional de Investigaciones Cient\'ificas y T\'ecnicas,
Rivadavia 1917, Buenos Aires, Argentina\\
$^{7}$University of Texas at Austin, McDonald
Observatory, 1 University Station, C1402, Austin, TX 78712-0259, USA\\
$^{8}$Shanghai Astronomical Observatory, Nandan Road
80, Shanghai 200030, China\\
\vspace*{-0.5cm}}

\begin{document}

\date{Accepted ???. Received ???; in original form ???}

\pagerange{\pageref{firstpage}--\pageref{lastpage}} \pubyear{2010}

\maketitle

\label{firstpage}

\begin{abstract}
We determine the intrinsic shapes and orientations of
$27,450$ type I and II active galactic nucleus (AGN) galaxies in the spectroscopic sample of the 
Sloan Digital Sky Survey Data Release 7, 
by studying the distribution of projected axis ratios of AGN hosts.
Our aim is to study 
possible alignments between the AGN and host galaxy
systems (e.g. the accretion disc and the galaxy angular momentum) and the effect of dust obscuration geometry on the AGN type.  
We define control samples of non-AGN galaxies
that mimic the morphology, colour, luminosity and concentration distributions of the AGN population, 
taking into account the effects of dust extinction and reddening. 
Assuming that AGN galaxies have the same
underlying three-dimensional shape distribution as their corresponding control
samples, {we find that the spiral and elliptical type I 
AGN populations are strongly {skewed} toward face-on galaxies, while ellipticals
and spirals type II AGN are {skewed} toward edge-on orientations.}
These findings rule out random orientations for AGN hosts at high confidence 
for type I spirals ($\delta \chi^2\approx 230$) and type II 
ellipticals ($\delta \chi^2\approx 15$), while the signal for type I ellipticals and type II
spirals is weaker ($\delta \chi^2\approx3$ and $\delta \chi^2\approx6$,
respectively).  {We obtain a 
much stronger tendency for the type II spirals to be edge-on }
when just high $[\rm OIII]$ equivalent width (EW) AGN are considered, suggesting 
that $>20$\% of low $[\rm OIII]$ EW edge-on type II AGN may
be missing from the optical sample. 
Galactic dust absorption of the broad-line region alone cannot explain the
observed inclination angle {and projected axis ratio} distributions of type I and II Seyfert types, 
implying that obscuration by a small-scale circumnuclear torus is necessary. 
These results favour a scenario in which
the angular momentum of the material which feeds the black hole retains a memory of its
original gas source {at least to some small}, non-negligible degree.
\end{abstract}


\section{Introduction}\label{Introsec}

Galaxies with Active Galactic Nuclei (AGN) have become a major focus
in extragalactic studies due to their role in galaxy formation scenarios
favoured today. Feedback from the AGN, and its effects
on gas infall and star formation are invoked to explain
observed trends on galaxy star formation rates, luminosities and colours
(see for instance,
\citealt{Bower06}; \citealt{Croton06}; \citealt{Cattaneo06};
\citealt{Sijacki07}; \citealt{Hopkins07}; \citealt{Marulli08}; \citealt{Lagos08};
\citealt{Somerville08}).
However, many aspects of the relationship between AGN and their hosts remain
unclear, such as the relation between the angular momentum of the
AGN system 
(i.e. accretion disc and black hole spin) and the galaxy kinematics. 
A level of coherence between the two could influence the development of the 
black hole (BH) spin \citep{Lagos09} which regulates the mass-to-energy conversion 
in radiatively efficient accretion phases (e.g. \citealt{Rawlings91}; \citealt{Marconi04}), 
and which has been postulated as key to the 
process of jet production (e.g. \citealt{Sikora07}); therefore, this could
directly
affect the energetic capacity of a BH to produce effective feedback.

A way to study galaxy-AGN orientations, 
for example in weak objects such as
radio-quiet AGN (i.e. Seyfert galaxies, \citealt{Osterbrock84}), 
is via the measurement of the inclination angles relative to the line of sight of AGN host galaxies,
since narrow (Seyfert II) and broad line (Seyfert I) AGN are thought to be the 
same physical phenomenon seen under different orientations
(see \citealt{Antonucci93} for a review). This is called the ``Unified
AGN model'' (e.g. \citealt{barthel89}; \citealt{madau94}; \citealt{Urry95}; \citealt{gunn99}) in which the emission lines come from the AGN system
composed of a BH surrounded by an accretion disc and an obscuring torus.
The accretion disc is responsible for the ionizing spectrum of the AGN, and the
broad line region (BLR) gas is distributed on similar scales; 
the direct observation of these regions results in a Seyfert I galaxy. The
opaque torus absorbs the optical emission from the 
accretion disc and the BLR along certain lines of sight, giving rise to a Seyfert II galaxy. 

The main evidence in favour of the unified model comes from observations of 
broad emission lines in polarised light from Seyfert II galaxies (e.g. \citealt{Antonucci85};
\citealt{Miller90}; \citealt{Zakamska05}; \citealt{Zhang08}; \citealt{Borguet08}),
and is further supported by e.g. measurements of similar BH {accretion rates} in
the two AGN types (\citealt{Netzer09}; \citealt{Hatziminaoglou09}).
However, the unified model does not imply a
relation between the inclination of the host galaxy and that of 
the obscuring torus (which may be linked 
with the BH spin; \citealt{Volonteri07},
\citealt{King08}; \citealt{Lagos09}), leaving open the question of the existence of such a relation.
{Indications of a relation between the kinematic
properties of the outer and inner galactic components 
have been recently found in the SAURON survey (\citealt{Bacon01}; \citealt{deZeeuw02})
by \citet{Dumas07} in AGN Seyfert galaxies. These show differences
in the position angle of outer and inner components 
typically of $\approx 30^o$, thus these two components are not randomly
oriented with respect to one other.
In addition, type II AGN spirals appear to be more elongated in optical bands than
are non-active spiral galaxies 
\citep{Shen10}, possibly indicating a tendency towards edge-on orientations in 
this type of AGN galaxies, which within the unified AGN model would point to a connection with the torus
orientation. On the other hand, \citet{Krajnovic08} found kinematically
decoupled cores (i.e. central regions)
in $30$\% of a small sample of elliptical galaxies observed as part of the SAURON project. Thus, the evidence
for alignment
between the galaxy as a whole and the AGN is not conclusive.}

Most of the studies in the literature on the 
orientations of AGN have been done using powerful radio
objects, by studying the level of alignment of relativistic jets
and their hosts. Since the jet direction is set by
the angular momentum of the accretion disc (e.g.
\citealt{Lynden-Bell06}) and/or the BH spin (e.g. \citealt{Blandford77};
\citealt{Fragile07}; \citealt{Barausse09}), it can be directly compared with the
orientation of the host galaxy. {Several papers (e.g.
\citealt{Kinney00}; \citealt{Schmitt02}; \citealt{Greenhill09}) 
have found that radio and optical position angles are uncorrelated, although their
samples contain $<100$ objects, meaning that the statistical 
precision of these studies is limited. Indeed, 
\citet{Battye09} found a significant level of alignment
between the minor axes of a sample of $6,053$
elliptical galaxies and the position angles of relativistic jets 
emitted from their central AGN.
However, such an apparent alignment could be difficult 
to distinguish from the ``alignment effect" 
(\citealt{McCarthy87}; \citealt{Chambers87}; \citealt{McNamara96a} and 1996b) in bright radio sources,  
in which the jet triggers star formation (e.g. \citealt{deYoung95}; \citealt{Blundell99}) or where 
the blue component is simply scattered AGN light \citep{McCarthy93}.

The BH spin development is intimately linked with the growth of the BH itself. 
This means that its value and its level of alignment 
with the galactic components give important clues about the way in which the accretion proceeds 
(e.g. \citealt{Moderski98}; \citealt{Hughes03}; \citealt{Shapiro05}).
The spin of a BH depends on whether it gained most of its mass via mergers with
other BHs or via accretion
(e.g. \citealt{Volonteri05}; \citealt{Berti08}; \citealt{Lagos09}).
{\citet{Volonteri05} found that gas accretion affects the spin evolution more
than do mergers between BHs, since binary coalescences alone do not lead to a
systematic spin-up or -down in time, while a prolonged accretion period 
efficiently spins BHs up to the
maximum value. On the other hand, \citet{King06}
suggested that the accretion proceeds via a series of short episodes as 
small amounts of gas with randomly oriented angular momenta fall in. 
In contrast with the \citet{Volonteri05} model, in this scenario BHs 
can achieve only low spin values, \^a$= c J_{\rm BH}/G M_{\rm BH}^{2} \lesssim
0.1-0.3$ through gas accretion 
\citep{King08} and relatively high spin values (i.e. \^a$\lesssim 0.7$; \citealt{Berti08}) can only be
achieved through the effects of galaxy and BH mergers \citep{Fanidakis10}.}
X-ray observations reveal relativistically broadened Fe-K$\alpha$ fluorescence
lines in several AGN, 
indicative of \^a$> 0.9$ (\citealt{Iwasawa96}; \citealt{Fabian02}; \citealt{Reynolds03};
\citealt{Brenneman06}). {Another approach to constrain BH spins is via
quasar demographics.
The observed light from quasars is directly proportional to the rate
at which supermassive black holes are accreting material, thus a
comparison of the quasar luminosity function with the present-day BH
mass function constrains the radiative efficiency \citep{Soltan82}, which
in turn depends on BH spin. Analyses based on this approach have come
to different conclusions: \citet{Elvis02},
\citet{Cao08}, \citet{Yu08} reported large energy conversion efficiencies
($>13$\%), indicative of large spins, while \citealt{Yu02}, \citealt{Marconi04},
\citet{Martinez-Sansigre09}, reported efficiencies as low as $7$\%, indicative
of slowly rotating BHs. This may depend on the redshift,
given that the typical accretion rates of relatively massive BHs are likely to 
be much larger at earlier epochs (e.g. \citealt{Wang09}).}

Lagos, Padilla \& Cora (2009, LPC09 hereafter) used the 
$\alpha$-disk model \citep{Shakura73} to follow the BH spin development in
the semi-analytic model of galaxy formation of \citet{Lagos08} within a
cosmological framework and showed that, in the hypothetical case in which
accretion discs preserve the angular momentum direction of the bulk of the cold
gas involved in star formation, massive galaxies
should host BHs with spin values close to unity, regardless of additional
physical effects such as accretion disc warps \citep{King05} or 
fragmentation \citep{King06}. We refer to this scenario, in which the BH is spun up smoothly via
steady accretion of gas that conserves the angular momentum direction of the original
gas source, as the coherent model. {Using the same semi-analytic model, the
authors tested the effect of 
assigning random orientations to the incoming material and show that high BH spin values 
can only be obtained by aligning the inner regions of the 
accretion disk with the BH spin (i.e. disc warps; \citealt{King06}); this case
will be referred to as the chaotic model.}

In this work we focus on the
chaotic/coherent accretion dichotomy, by studying 
intrinsic shapes and inclinations of host galaxies of AGN with obscured and 
unobscured BLR { in the Sloan Digital Sky Survey (SDSS, \citealt{York00})
Data Release 7 (DR7, \citealt{Abazajian09})},
to determine the degree of alignment between galaxy
and accretion discs (or tori).
{Throughout the paper the term AGN will refer
only to Seyfert galaxies.}
We will assume that the torus and the accretion disc are aligned, with the exception of
a possible warp in the accretion disc in the vicinity of the central super-massive BH (\citealt{King05}) 
as seen in AGN masers (e.g. \citealt{Zhang06}).
We compare the
results obtained with predictions of the LPC09 model in the coherent and chaotic 
scenarios to constrain the physics of gas inflow.
 
{This paper is organised as follows. In Section~$2$, we
describe the sample and the selection of control samples matched to the properties of AGN galaxies. 
We characterise the three-dimensional intrinsic shapes of these samples following
\citet{Padilla08}. 
In Section~$3$ we study the orientations of AGN
host galaxies, systematic effects in the AGN selection, and the origin of 
biases toward face- or edge-on orientations by analysing the effect of galactic
disc and nuclear torus obscuration. 
In Section~$4$ we compare the observed inclination angle distributions 
with theoretical predictions for the chaotic and coherent models of LPC09.
We discuss and summarise our main results and 
their implications for AGN models in Section~$5$ and Section~$6$,
respectively.}
Throughout this paper we assume a cosmological model characterised by
matter and dark-energy density parameters $\Omega_m=0.27$, $\Omega_{\Lambda}=0.73$ and 
a Hubble constant $H_0=73\,\rm km\,s^{-1}\, Mpc^{-1}$.  These parameters are consistent with the 
results from the Wilkinson Microwave Anisotropy probe (WMAP, \citealt{hinshaw09}, \citealt{dunkley09},
\citealt{sanchez06}).

\section{The intrinsic shapes of AGN host galaxies in the SDSS}

In this Section we 
characterise the three-dimensional intrinsic shapes of the hosts of AGN. Our AGN sample 
was identified using the method of \citet{Hao05a} 
from the SDSS DR7 spectra. We assume that the AGN have the 
same intrinsic three-dimensional shapes as well-defined control samples of non-AGN galaxies. 
{Galaxies in the control samples should be free of inclination biases, 
and therefore their observed projected shapes 
will only reflect their intrinsic three-dimensional shapes, plus the effect of extinction.} 
Comparison of the axis ratio distribution of AGN galaxies with those of the control sample allows us to make 
inferences on the orientation distribution of the AGN.

\subsection{The SDSS DR7 AGN sample}

The AGN data used in this work consists of $6,153$ type I and $21,297$
type II AGN identified by their emission-line properties from about $698,000$ galaxy spectra 
(the MAIN survey; \citealt{Strauss02}) 
in the SDSS DR7 using the methods of \citet{Hao05a}. 
{In order to exclude objects with low $\rm S/N$ emission lines, Hao et al.
reject from the sample galaxies whose $\rm H\alpha$ rest-frame equivalent width (EW) is $<3 \AA$.
We exclude type III AGN (\citealt{Kauffmann03}), 
which lie between the star-forming and AGN loci \citep{Kewley01} in the \citet{Veilleux87} line ratio diagram,  
since much of their emission could come from star-formation activity. 
Hao et al. fit for and subtract the stellar continuum before
measuring the strength and widths of emission lines.
H$\alpha$ and H$\beta$ are fit to two Gaussians; those objects with a significant
broad component ($FWHM>1200$km/s) are termed type I, while those objects lying
beyond the star-forming line ratio limits of \citet{Kewley01} are termed type
II. 

We separate the AGNs further by morphology, using the fraction of $r$-band light 
which fits a de Vaucouleurs profile, $f_{\rm dev}$ (from model fits 
available from the SDSS pipeline, \citealt{Abazajian04}).
Early-type galaxies are selected by requiring $f_{\rm dev} \ge
0.9$, and we classify all other galaxies as spirals. We therefore define
four AGN samples, selected by AGN type and morphological host galaxy type.

We apply a dust extinction and reddening
correction to luminosities and colours, respectively, of spirals using the model by Padilla \& Strauss
(2008, PS08 hereafter), where it is assumed that the amount of reddening is 
proportional to the path length of the light through the galaxy. 
The extinction is a function of the inclination angle $\theta$ for any given galaxy,
the mean galaxy height to diameter ratio, $\gamma=1-C/B$ (where galaxies
are modelled as triaxial ellipsoids of major, middle and minor axis $A$, $B$, and
$C$, respectively), and the edge-on extinction, $E_{0}$ (in
magnitudes). {Here, a perfect edge-on galaxy will have
$\theta=90^o$, while a face-on galaxy has $\theta=0^o$.}
For each spiral galaxy we obtain $E_{0}$ and $\gamma$ from Table~$4$ of PS08 (interpolating
in absolute magnitude).
In turn, $\theta$ is determined from the observed projected axis ratio 
by interpolating in their Table~$8$.
The edge-on reddening in $r$ magnitudes, $R_0$, is related to $E_0$ via $E_{0}=f\, R_{0}$, where
$f=2.77$ is the reddening parameter.
PS08 showed that the model is not strongly dependent on $f$, allowing us to use
this simple approximation to correct both colours and luminosities.
The extinction affects the maximum volume, $V_{\rm max}$, out to which a galaxy
can be detected in the SDSS flux-limited catalogue.
We select galaxies with $M_{\rm r}-5 \rm log_{10}\,h < -17$ in order to
avoid very small values of $V_{\rm max}$, since fainter galaxies tend to
dominate the noise in the estimate of the distribution functions we consider below.

\subsection{Selection of Control Samples}

We select control samples matching the colours ($g-r$), 
luminosities ($M_r$) corrected for reddening and extinction, respectively, and 
concentrations\footnote{Concentration is defined as the ratio 
between the radii enclosing $90$\% and $50$\% of the \citet{Petrosian76} 
light, $r_{90}/r_{50}$.} of the AGN host samples.  All magnitudes shown are $k$-corrected to $z=0.1$. 
We then construct a grid in the parameter space defined by 
these three galaxy properties, and populate each cell with the AGN hosts weighted by $1/V_{\rm max}$.
The control samples are constructed by filling each grid cell with randomly chosen main sample 
galaxies from the SDSS so that $\Sigma\, V_{\rm max, Gals}^{-1} \approx 
N \times \Sigma\, V_{\rm max,AGN}^{-1}$; $N$ is chosen
so as to maximise the size of the control samples, with a maximum departure from the
target distribution in any grid cell of less than $20$\%.
$N$ is $5-20$ for our four AGN samples. Notice that
this selection takes into account any intrinsic correlation between these quantities in
the AGN sample. This ensures that control samples mimic the AGN hosts as closely as possible.
Table~\ref{nobjs} shows the number of galaxies in each AGN sample and their
corresponding control sample.

\begin{table}
\begin{center}
\caption{Number of galaxies in each AGN sample and their corresponding control
sample.}\label{nobjs}
\begin{tabular}{c c c}
\\[3pt]
\hline
\small{Sample} & AGN sample & control sample\\
\hline
Type I ellipticals & 2,712  & 50,537 \\
Type II ellipticals & 7,653  & 45,253 \\
Type I spirals & 3,360 & 57,010 \\
Type II spirals & 13,578 & 66,626  \\
\hline
\end{tabular}
\end{center}
\end{table}

\begin{figure*}
\begin{center}
\includegraphics[width=0.99\textwidth]{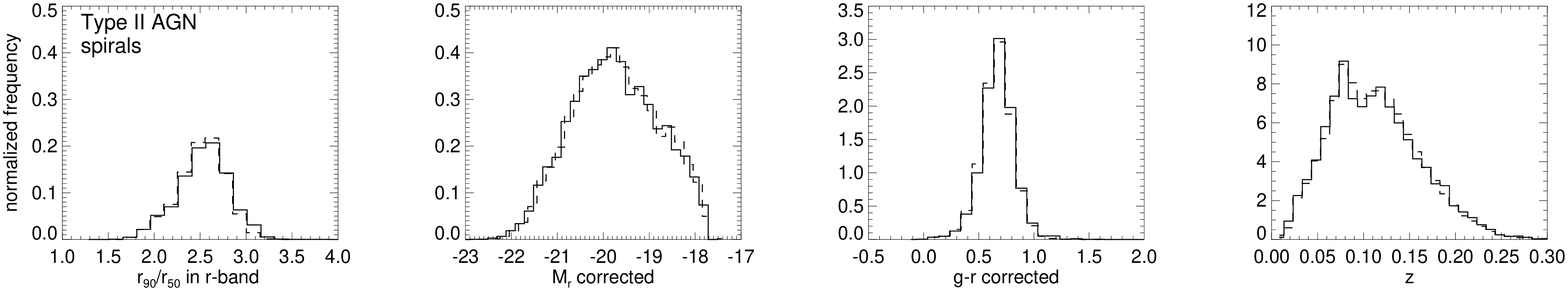}
\includegraphics[width=0.99\textwidth]{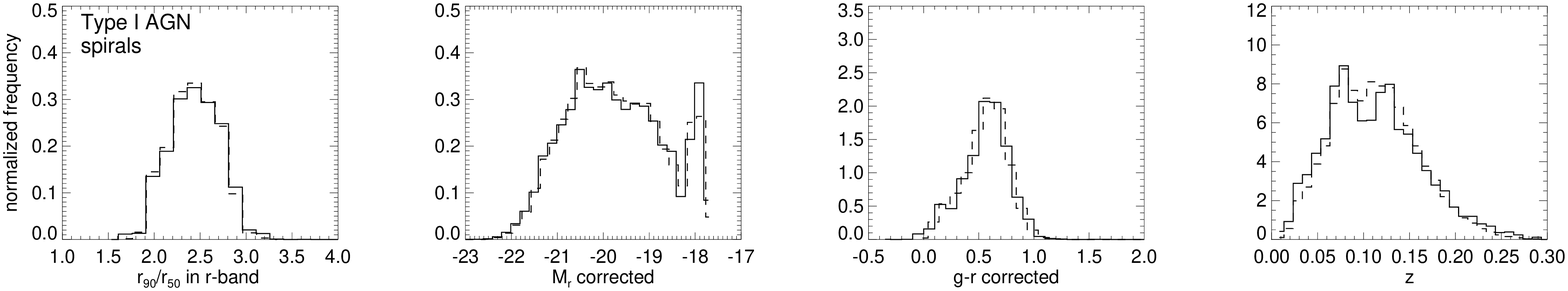}
\includegraphics[width=0.99\textwidth]{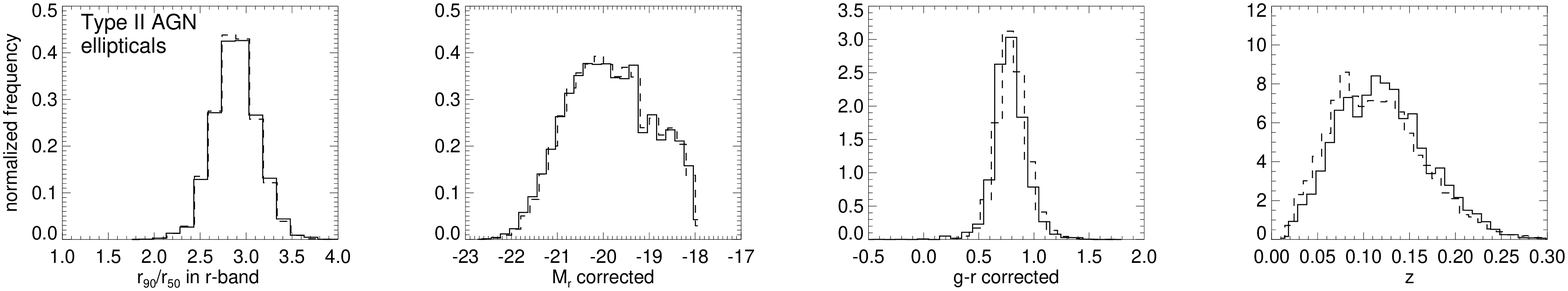}
\includegraphics[width=0.99\textwidth]{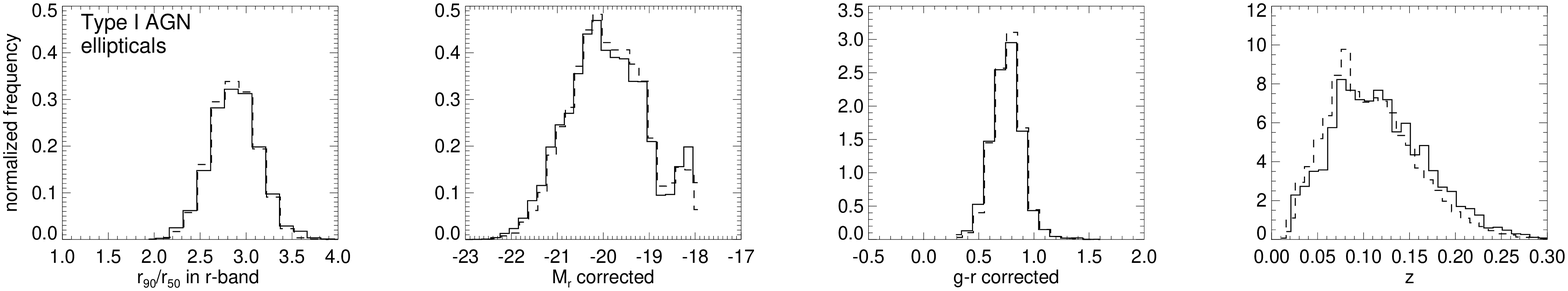}
\caption{\label{select1} $V_{\rm max}$-weighted distributions of $r_{90}/r_{50}$ (left panels), $M_{\rm
r}-5\rm log_{10}\, \rm h$ (middle-left panels), $g-r$ colour (middle-right), and redshift
distributions (right panels) for AGN (solid lines) and control 
galaxy samples (dashed lines), divided according to AGN and morphological type.  
Absolute magnitudes and colours were corrected for internal extinction and
reddening, respectively.}
\end{center}
\end{figure*}

Fig.~\ref{select1} shows the $r_{90}/r_{50}$ (left panel), $M_{\rm r}-5\rm
log_{10}\, \rm h$ (middle-left
panel) and  $g-r$
(middle-right panel) $V_{\rm max}$-weighted distributions of each of the AGN subsamples 
(solid lines) and their corresponding control
samples (dashed lines). 
The right panels show the redshift distributions of
both populations.  
The control sample distributions all follow the AGN
distributions very closely.
\citet{Hao05b} noted that sufficiently low luminosity AGN would not be selected
spectroscopically due to
(i) undetectable emission lines, (ii) higher noise in the measurements of
emission line ratios for weaker emission line strengths, or (iii) total galaxy fluxes 
(including the nuclear emission) below the spectroscopic sample limit. 
This could result in a difference in the redshift distribution of AGN and non-AGN galaxies, which
would cause a difference in the values of $V_{\rm max}$.
The observed agreement between the redshift distributions of AGN and non-AGN 
galaxies (right panel of Fig.~\ref{select1}) 
suggests that this is not a problem. We also find a broad agreement between the $M_{\rm r}-5\rm
log_{10}\, \rm h$ distributions of AGN and control samples for the spiral
population, and only a small offset at the flux limit ($m_r\approx17.5$) in
the elliptical population. We conclude that our $V_{\rm max}$ does not need extra corrections for these effects 
and is adequate for our analysis.

In Fig.~\ref{select1}, the host galaxy properties of the two AGN
types (for the same morphology) show very similar photometric properties, a 
result consistent with the AGN unified model, in 
which the difference between the types is due solely to orientation. 
Even though the non-obscured nuclear emission from type I AGN could affect the photometry of the
galaxies, the similar distributions
of luminosities and concentrations of type I and type II AGN indicate that such an effect is minor.
{In fact, type I AGN galaxies are only $\Delta(g-r)\simeq0.1$ redder than type II
AGN, smaller than the binsize used.} 
The colour distributions of spiral and elliptical AGN hosts differ only slightly; elliptical
hosts are on average only $\Delta(g-r)\simeq0.2$ redder than spiral hosts.
This small difference
is common in AGN hosts, as they often lie in the ``green valley" of the colour-magnitude
relation diagrams (see for instance \citealt{Padilla10}). 

We have thus constructed normal galaxy samples that reproduce the
colour-luminosity-concentration properties of AGN populations (selected by
AGN type and host morphology) which, unlike AGN of a given type, are expected to be randomly
oriented with respect to the plane of the sky. We now characterise the intrinsic typical shapes
and orientations of the selected galaxy control samples.

\subsection{The intrinsic shapes of AGN galaxies in the SDSS}

We measure the intrinsic galaxy shapes in each AGN sample 
by studying the projected $b/a$ distribution in the
$r$-band for the control samples. 

\begin{figure*}
\begin{center}
\includegraphics[width=0.4\textwidth]{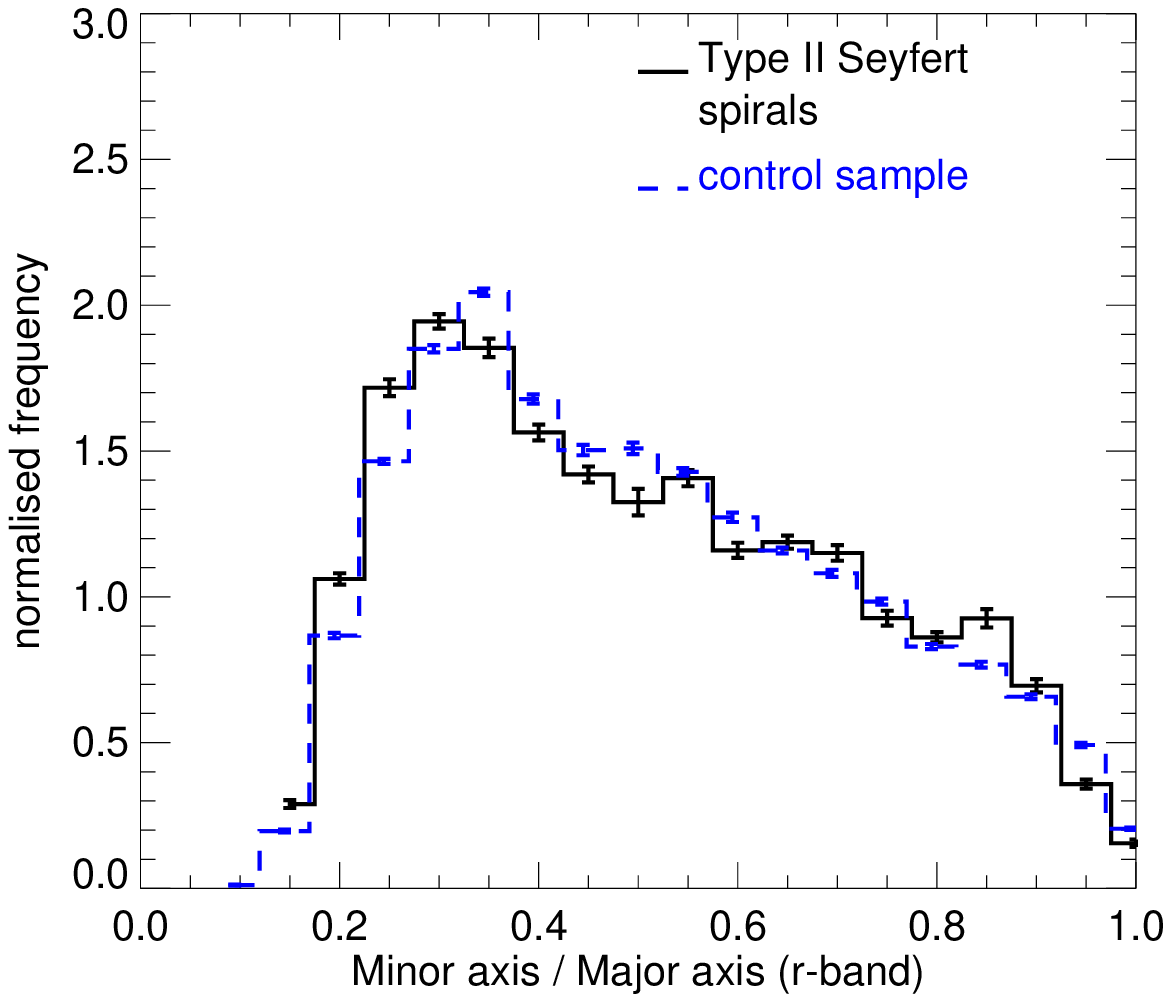}
\includegraphics[width=0.4\textwidth]{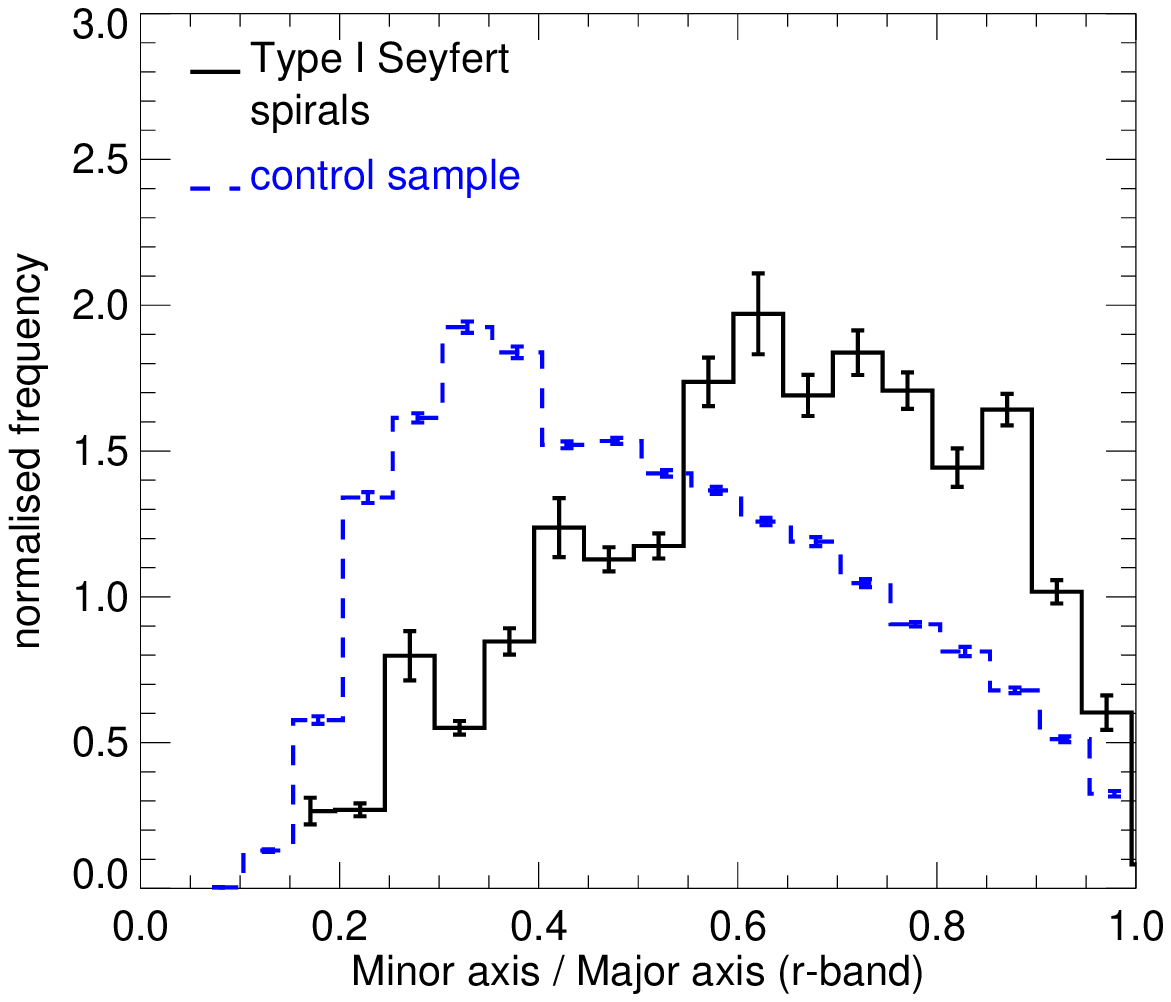}
\includegraphics[width=0.4\textwidth]{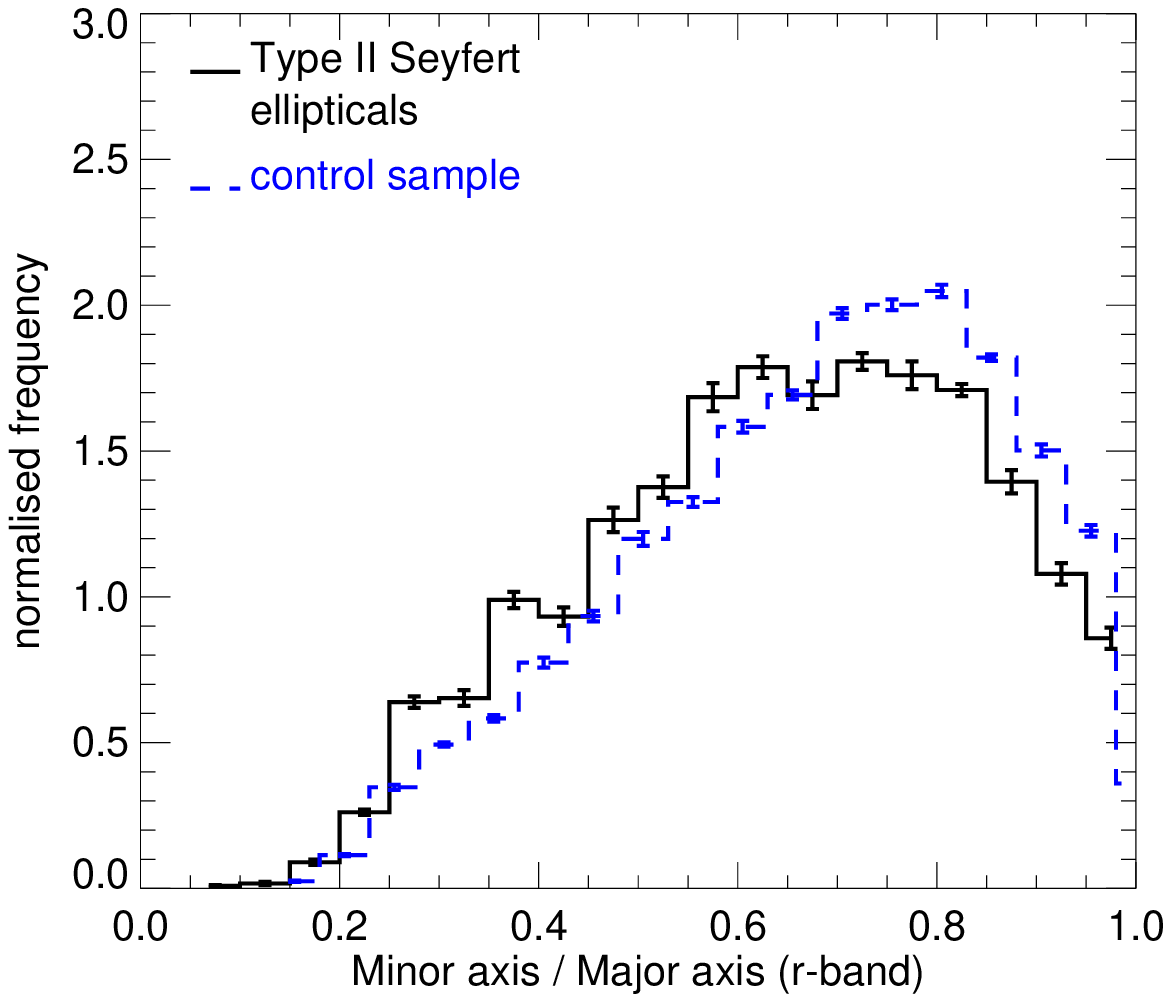}
\includegraphics[width=0.4\textwidth]{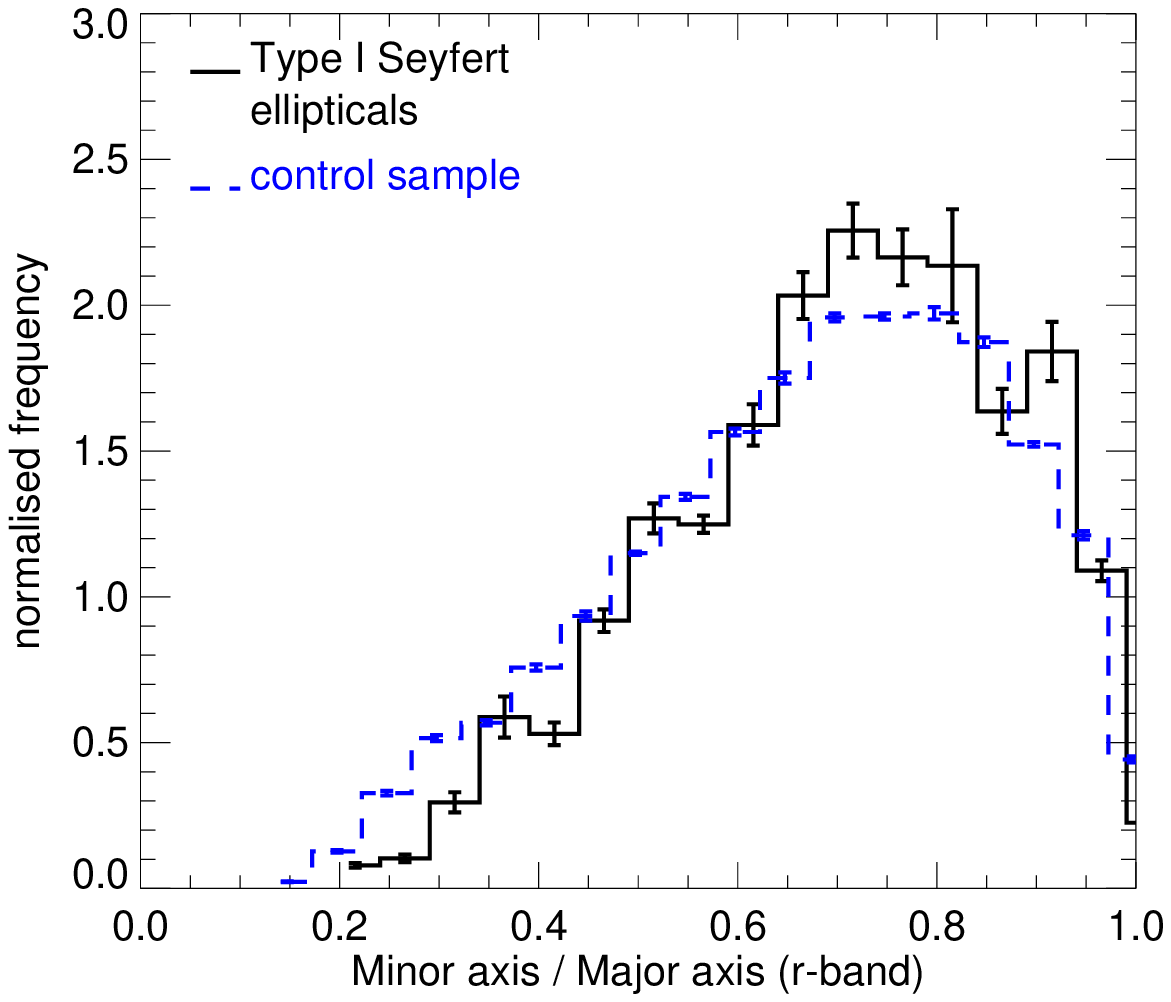}
\caption{\label{AGNIIEll} 
Distributions of projected axis ratio, $b/a$, for selected  spiral galaxies (left) and
elliptical galaxies (right) hosting a type II (left panels) and type I (right panels) AGN
(solid lines) and the corresponding control sample (matched to the distribution in
the colour-luminosity-concentration distributions that the AGN population follows; dashed lines). 
{All the distributions have been weighted by $1/V_{\rm max}$, corrected for
extinction. Errors were
calculated using the jackknife technique.}}
\end{center}
\end{figure*}

Fig.~\ref{AGNIIEll}
shows the distribution of minor to major projected axis ratios ($b/a$ from exponential
or de Vaucouleurs fits corrected for the effects of the PSF)
for spiral (top) and elliptical (bottom) type I (left panels) and type II
(right panels) AGN galaxies
(solid lines) and the corresponding control samples (dashed lines).
All the distributions shown are $V_{\rm max}$-weighted.
Errors are calculated using the jackknife technique. 
Both type II AGN ellipticals and spirals seem to be slightly more edge-on
than their control samples, 
which are assumed to cover the full range of
inclination angles. The shift in the $b/a$ distribution peak 
towards lower $b/a$ is about $\approx 0.05$ and $\approx 0.15$ for the spiral and elliptical type II
populations, respectively.
There are considerably larger differences between the
$b/a$ distributions of the spiral type I AGN population and its control
sample; the AGN population is {skewed} towards high
$b/a$ values, indicating inclination angles {closer to face-on orientations}. 
The control sample shows a peak in the distribution at $b/a \approx 0.35$, while the AGN
population peaks at $b/a\approx 0.65$. 
This suggests that the source of obscuration in type II objects has a large scale height,
and is preferentially aligned with the plane of the galaxy disc.

{At first glance, the larger inclination effect for type I AGN than that observed 
for type II AGN could be taken as a failure of the unified
AGN model, since it naively suggests that the relation between the host galaxy
and its AGN depends on the AGN type. However, a proper comparison between the two types of AGN
requires a more detailed analysis of relative inclination angles which we will
perform in Section~\ref{sec:inclinations}; another possibility that
we will also explore is that the samples of different types of AGN are subject to
different selection effects.}

\subsubsection{Characterisation of the intrinsic shapes}

We use the $b/a$ distributions to characterise the three-dimensional shapes of galaxies in
each control sample. A variety of papers (\citealt{Ryden04}, \citealt{Vincent05}, \citealt{Padilla08})
have shown that both elliptical and spiral galaxies have shapes consistent with
oblate spheroids.
In this work we model galaxies as
triaxial ellipsoids of major axis A, middle axis B and minor axis C 
parametrised by two axis ratios, $B/A$ and $C/B$. We follow the PS08 assumption 
that the distribution of $1-C/B$ follows 
a Gaussian with mean $\gamma$ and standard deviation
$\sigma_{\gamma}$, and the distribution of
$\epsilon=\ln(1-B/A)$ is Gaussian with mean $\mu$ and dispersion
$\sigma$. {Indeed \citet{Ryden04} found that the distributions of intrinsic axis ratios are consistent with the
Gaussian and log-normal shapes we are using.}
We use the PS08 model to fit the observed $b/a$ projected distribution in order to determine these 
parameters for each galaxy sample.
This approach also requires a model for internal 
extinction, as we described above. An important quantity is $\psi(\theta)$, the
ratio of the number of galaxies seen at inclination $\theta$ to
the number expected without extinction, 

\begin{equation}
\psi(\theta)=
\frac{\int_{-\infty}^{\infty}\hskip-.08cm\int_{-\infty}^{\infty}\hskip-.08cm
f_E(M)f_R({\cal C})\phi(M)\phi({\cal C})W\hskip-.08cm({\cal C},M){\rm
d}{\cal C}{\rm d}M}
{\int_{-\infty}^{\infty}\int_{-\infty}^{\infty}\phi(M)\phi({\cal C})W({\cal
C},M){\rm d}M{\rm d}{\cal C}},
\label{model}
\end{equation}

\noindent where $\phi(M)$ and $\phi(\cal C)$ are the uncorrected luminosity (here in the r-band) and
colour (here $g-r$) functions, respectively, for a given sample. 
The function $f_{\rm E} (M)=\phi_{E}(M)/
\phi (M)$ is the ratio between the number of observed extincted galaxies 
and intrinsic number of galaxies at a given luminosity, 
where the subindex $E$ indicates that it refers to the extincted luminosity function.
Similarly, $f_{R} (\cal C)$ is the ratio between the underlying and reddened
distributions of galaxy colours. $W$ contains the correlation between
$g-r$ and $M_{r}$, which we assume to be Gaussian with mean and dispersion as
measured directly from the data. The function $\psi (\theta)$ is thus completely
characterised by the dust-corrected luminosity and colour functions, 
and by the parameters of the dust and shape model, namely $E_{0}$,
$\mu$, $\sigma$, $\gamma$ and $\sigma_{\gamma}$.
A grid of parameters is constructed, and for each
grid point, $p_i$, we obtain a model $N_{\rm model}(b/a,\{p\}_i)$ projected axis
ratio distribution, which is compared to the observed distribution;

\begin{equation}
\chi^2(\{p\}_i)=\sum_{b/a\,\rm bins}\left( 
\frac{N_{\rm model}(b/a,\{p\}_i)-N(b/a)}{\sigma_{\rm jack-knife}(b/a)}\right)^2.
\label{chi}
\end{equation}

\noindent The best-fitting parameters correspond to the minimum value of $\chi^2$
throughout the parameter grid. Following PS08, we assume $E_0=0$ for elliptical galaxies.

\begin{table*}
\begin{center}
\caption{Best-fitting model for each of the four control galaxy samples
selected to mimic the colour, luminosity and concentration distributions
of spirals and
elliptical Type I and II AGN. The last column shows the likelihood associated
with the parameter set.}\label{ParametersPS08}
\begin{tabular}{c c c c c c c}
\\[3pt]
\hline

\small{Sample} & \small{$E_{0}$} & \small{$\mu$} & \small{$\sigma$} &
\small{$\gamma$} & \small{$\sigma_{\gamma}$} & \small{$P_{\rm max}$}\\
\hline
Type I ellipticals & $\equiv 0.0$  & $-0.9$ $\pm$ 0.5 & 2.3 $\pm$ 0.6 & 0.45 $\pm$ 0.03 &
0.21
$\pm$ 0.04 & 0.86   \\
Type II ellipticals & $\equiv 0.0$  & $-1.35$ $\pm$ 0.4 & 1.7 $\pm$ 0.5 & 0.45 $\pm$ 0.03 &
0.23
$\pm$ 0.04 & 0.88 \\
Type I spirals & 0.3 $\pm$ 0.3  & $-0.85$ $\pm$ 0.35 & 1.7 $\pm$ 0.2 & 0.75 $\pm$
0.02
& 0.07
$\pm$ 0.03 & 0.25 \\
Type II spirals & 0.3 $\pm$ 0.3  & $-0.25$ $\pm$ 0.3 & 2.2 $\pm$ 0.2 & 0.75 $\pm$
0.02
& 0.04
$\pm$ 0.02 & 0.33  \\
\hline
\end{tabular}
\end{center}
\end{table*}

Fig.~\ref{ModelAB}
shows the best-fitting model compared to the observed distributions for each
control galaxy sample.
Symbols represent the 
observed SDSS galaxy $b/a$ distributions, and solid lines are the best-fitting models. 
The best fit parameters for each population are given in
Table~\ref{ParametersPS08}. The likelihood, $P_{\rm max}$, that corresponds to the minimum
$\chi^2$ value quantifies whether our model is a good fit to the data, and is given for each case. 
The model axis distributions follow the observed ones well. The values of $\mu$ and $\gamma$ for both elliptical galaxy
populations are consistent with oblate spheroids, in agreement with previous
analyses (e.g. \citealt{Ryden04}; \citealt{Vincent05}; \citealt{Padilla08}; \citealt{Battye09}). 
Moreover, control samples of both morphological types 
are characterised by shapes consistent (within the errors) with those of the
full galaxy population as measured by PS08
using the SDSS DR6 \citep{Adelman-McCarthy06}.  
The measured distributions, as well as the best fit values of $E_0$, $\gamma$ and $\sigma_{\gamma}$,
are similar for the type I and II AGN spirals; $\mu$ and $\sigma$ show differences which could
be used to argue against the unified model, but these are only marginally 
significant (slightly above one standard deviation; {errorbars are given in Table~\ref{ParametersPS08}}).
These slight differences could be due to different selection effects acting on each AGN type, an issue
that we will explore further in Section \ref{ssec:ew}.

\begin{figure}
\begin{center}
\includegraphics[width=0.5\textwidth]{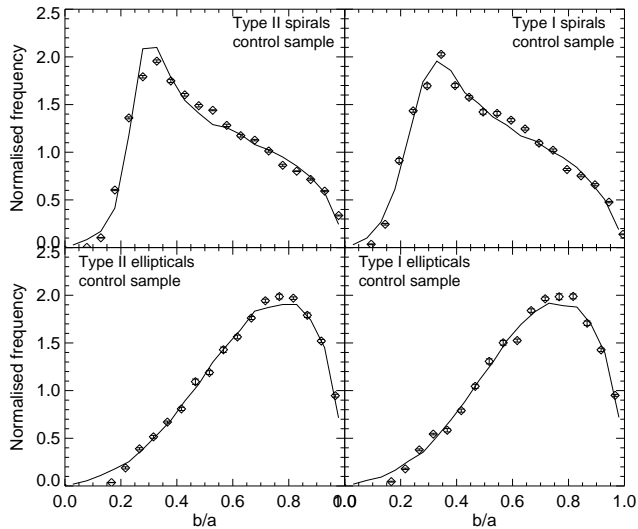}
\caption{\label{ModelAB}
Distributions of projected axis ratios, $(b/a)$, weighted by $1/V_{\rm max}$ 
for the control samples of galaxies selected to
populate the same colour-luminosity-concentration space as our four classes of AGN.
Symbols correspond to the control samples selected using the SDSS DR7
spectroscopic sample, and lines are the
best-fitting models for the parameters shown in  Table~\ref{ParametersPS08}.
}
\end{center}
\end{figure}

In order to analyse possible parameter degeneracies we study the
probability contours (calculated from $\chi^2$)
in the $\mu-\gamma$ and $E_0-\mu$ spaces. 
Fig.~\ref{contout2} shows $\mu-\gamma$
probability contours for the control samples of each AGN sample (as labelled). The white region
surrounding the contours represents probabilities $< 0.01$. 
The values of $\mu$ and $\gamma$ are very well constrained, particularly for
spiral galaxies, which show a very well defined region of high likelihood,
while for elliptical galaxies the contours tend to
be slightly elongated and tilted, allowing a larger range of possible parameter values in $\mu$.
This may be due to the lower sensitivity of the method to large negative values of 
$\mu$, which are associated with very round shapes.

\begin{figure*}
\begin{center}
\includegraphics[width=0.43\textwidth]{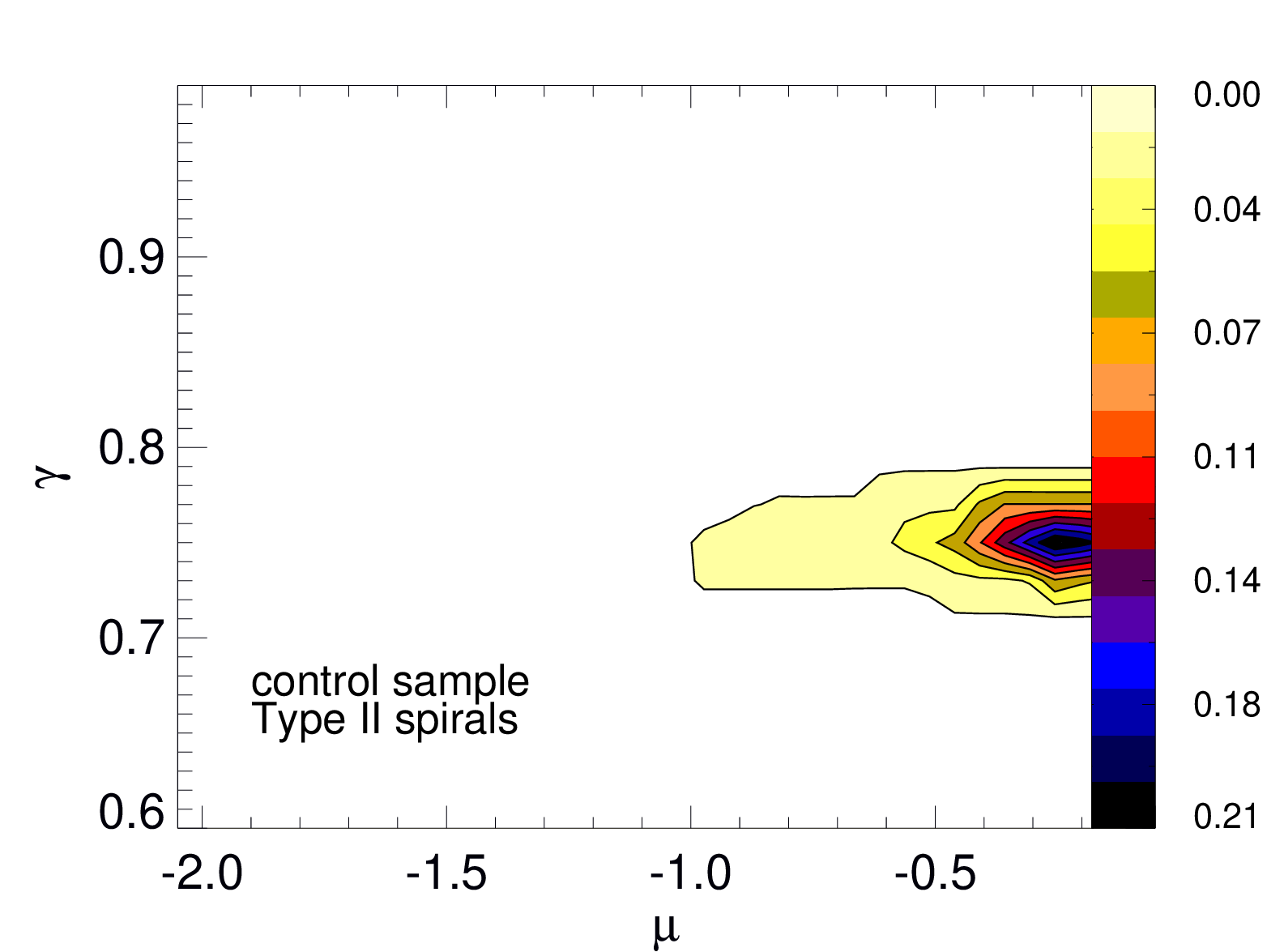}
\includegraphics[width=0.43\textwidth]{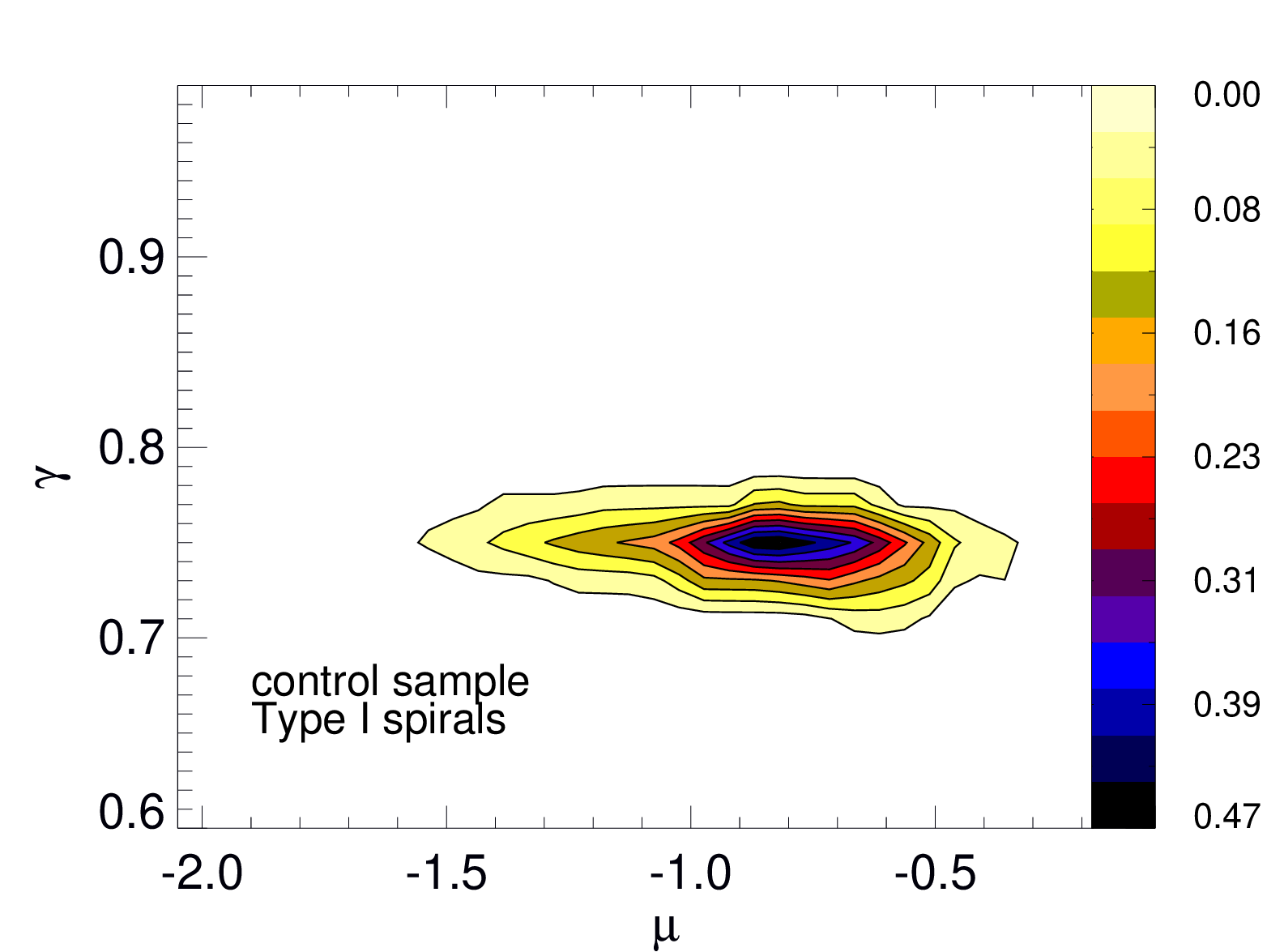}
\includegraphics[width=0.43\textwidth]{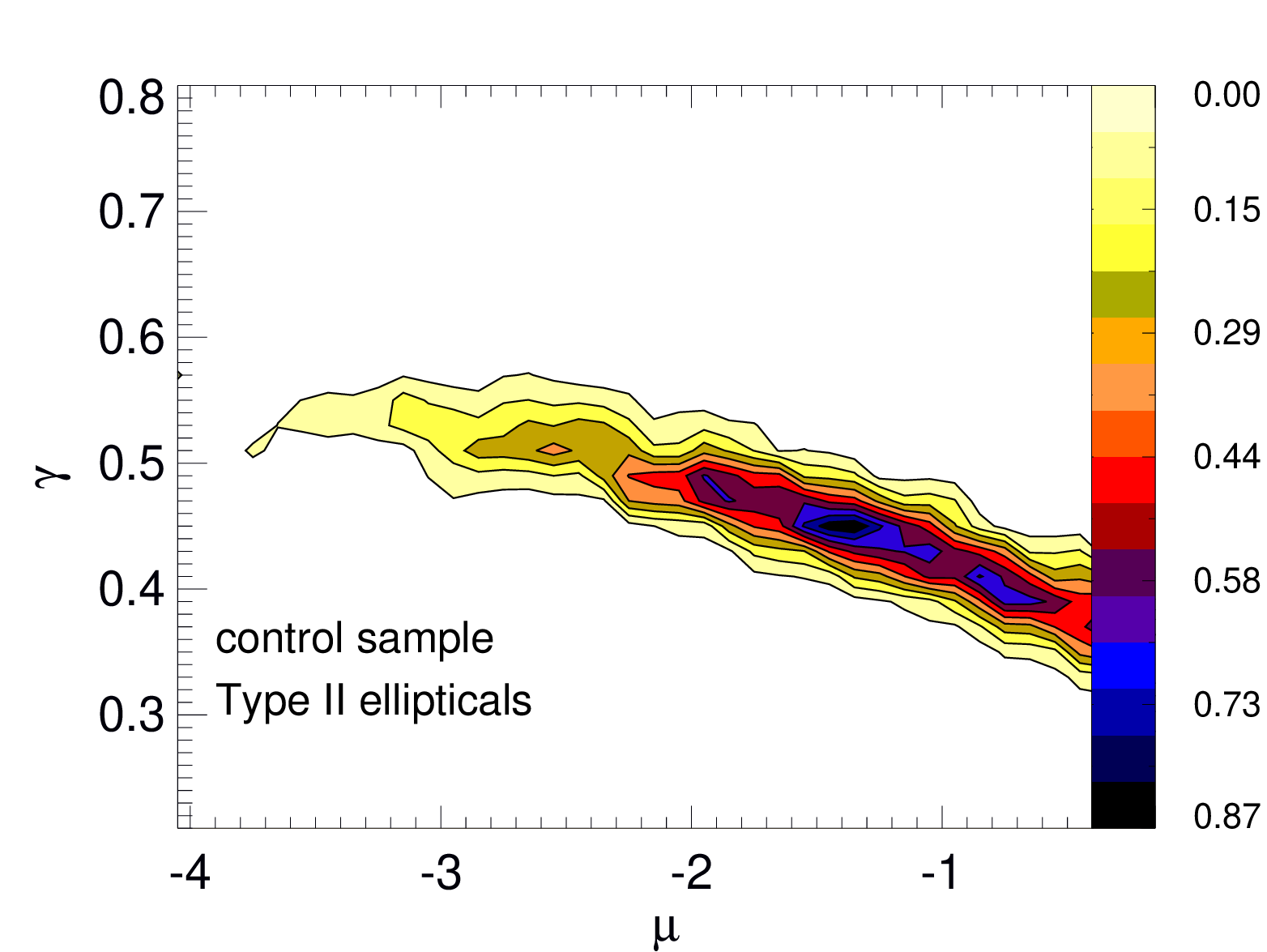}
\includegraphics[width=0.43\textwidth]{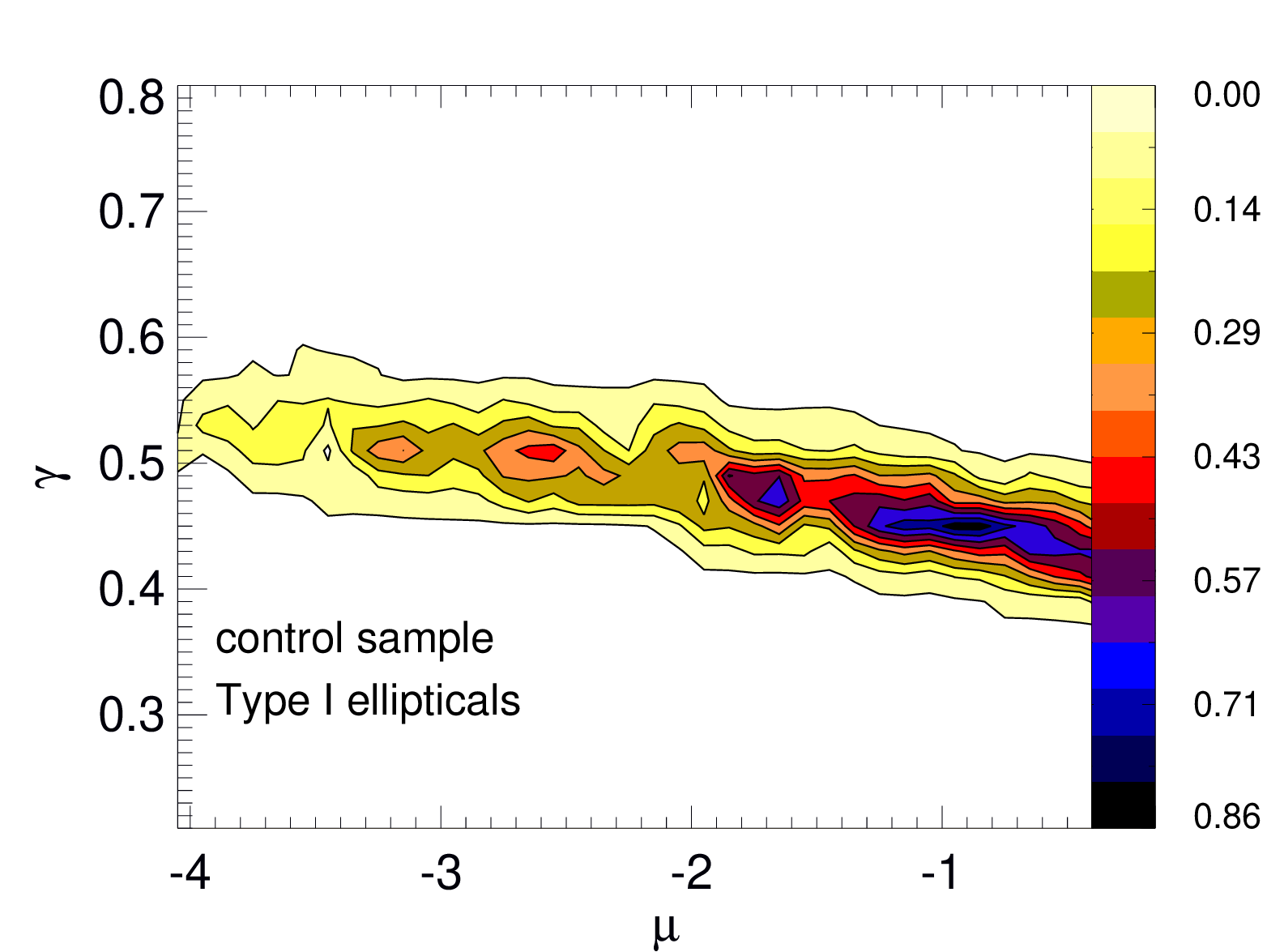}
\caption{\label{contout2} Probability contours in the 
$\mu$-$\gamma$ plane, for galaxy samples selected to mimic type II (upper left) and type I
(upper right) spiral AGN, and type II (lower left) and type I (lower
right) elliptical AGN. {\rm The colour bar at the right of each panel shows the range of
likelihood plotted.}}
\end{center}
\end{figure*}

We also inspected the $\mu-\sigma$ probability contours for all control
samples (not shown) and find that these two parameters are slightly more
degenerate than the $\mu-\gamma$ plane.
The quantity $\sigma$ smooths the $N_{\rm
model}(b/a)$ distributions, and therefore, the relatively large uncertainty in this parameter simply
indicates that it does not have an important effect on the shape of the
modelled $b/a$ axis ratio distributions. 

Fig.~\ref{contout1} shows the probability contours in the $E_0-\mu$ space for
the control samples of the type II (top panel) and type I (bottom panel) AGN
spiral galaxy samples ({ellipticals are assumed to have $E_0=0$}). The extinction is
only marginally detected in both cases.  Dust strongly determines the likelihood that a galaxy will
enter the sample at a given viewing angle, which consequently affects the observed
axis ratio distribution. {Note that $\mu$ is very insensitive to $E_0$,
except for extreme values of $\mu$ (i.e. $B/A \ll 1$). This means that when 
galaxies are intrinsically elongated (independent of their inclination angle), a
large extinction is needed in order to broaden the projected $b/a$ distribution
and move its peak towards larger $b/a$.}

\begin{figure}
\begin{center}
\includegraphics[width=0.43\textwidth]{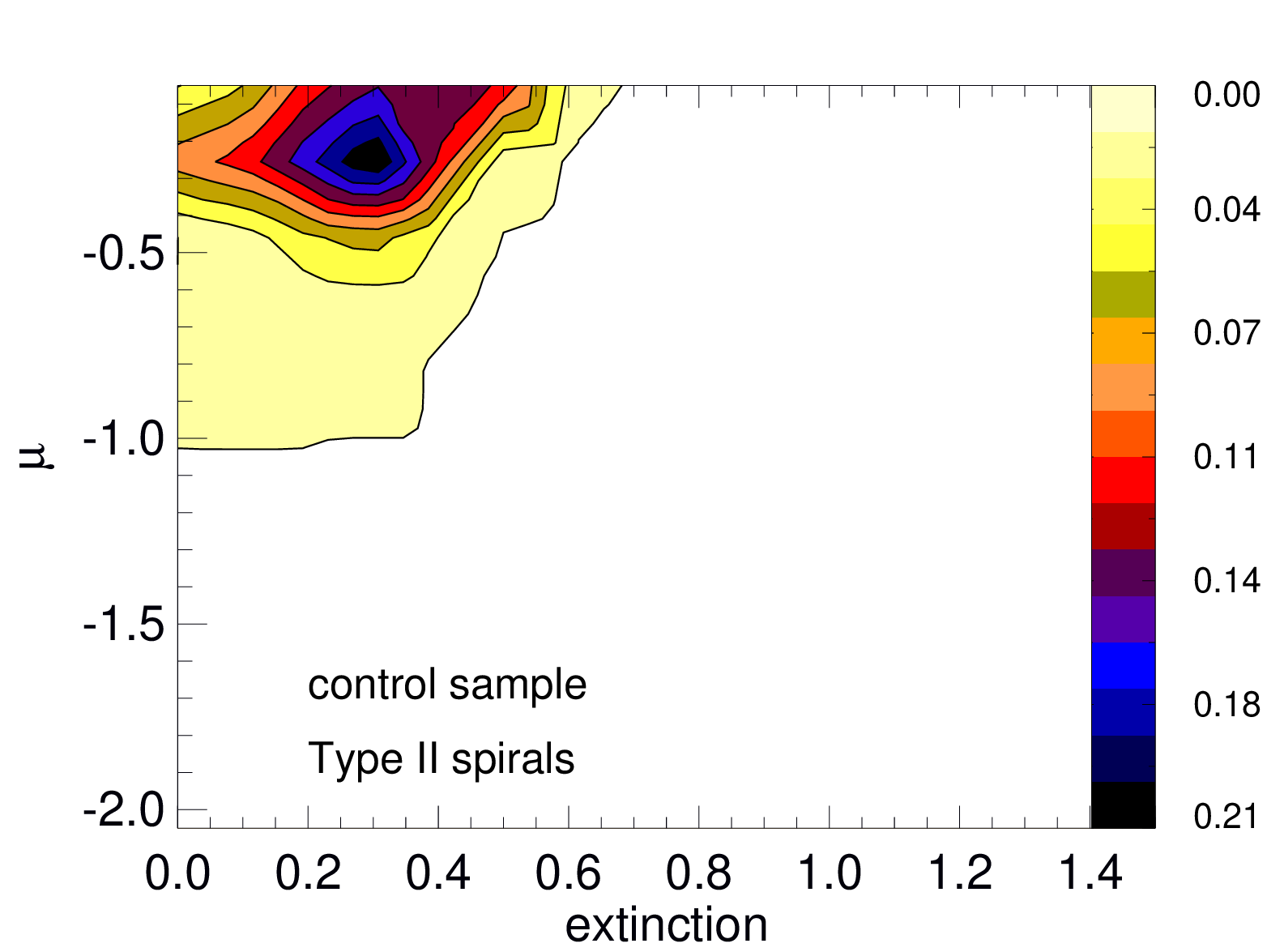}
\includegraphics[width=0.43\textwidth]{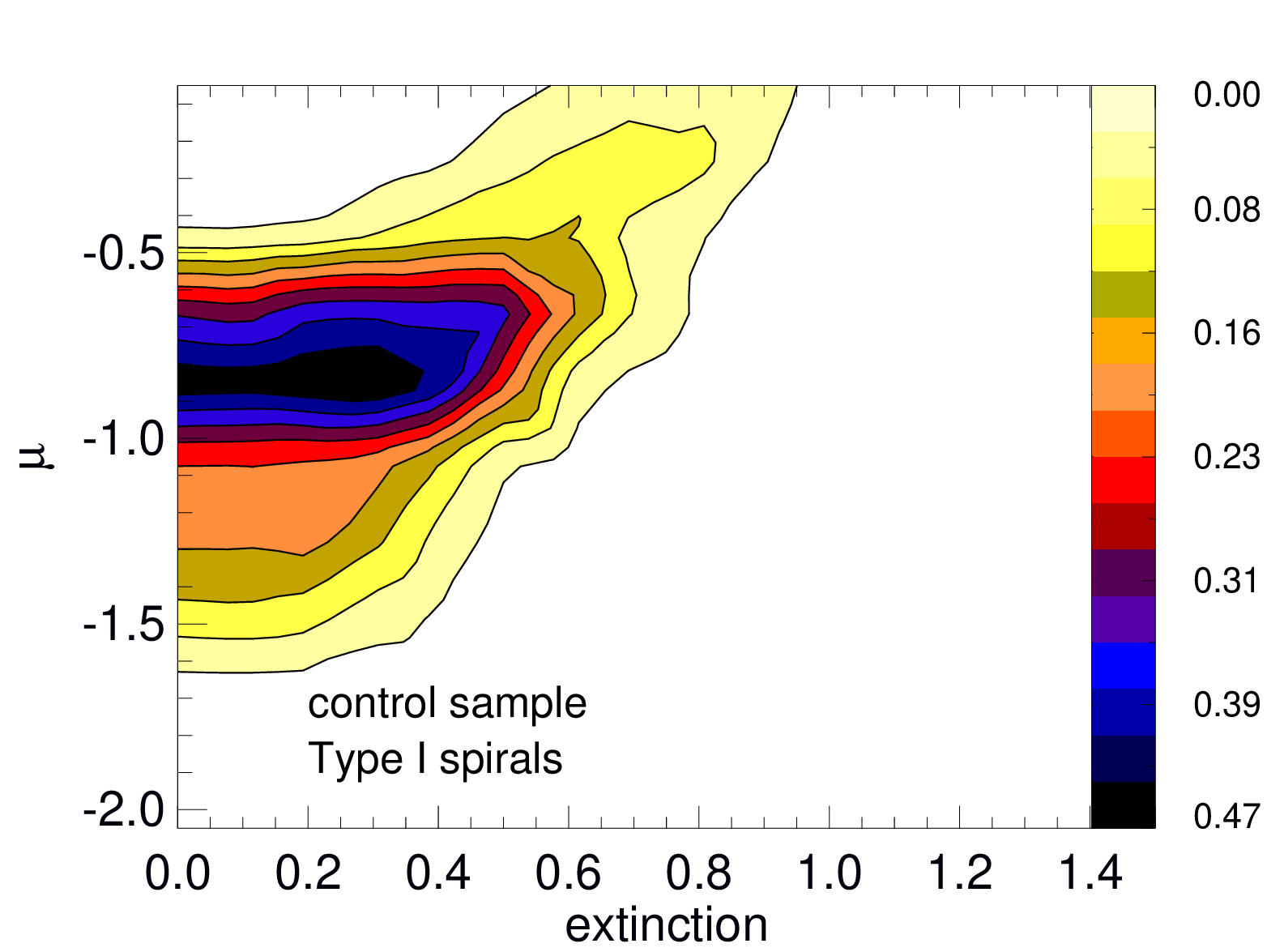}
\caption{\label{contout1} Probability contours in the $E_{0}$-
$\mu$ space, for the control samples of AGN spiral galaxy samples (as labelled).  
The colour-scale is shown on the right of each panel.}
\end{center}
\end{figure}

{The
modelled $b/a$ distributions do not show significant changes within the errors
calculated for each parameter, implying that the confidence regions are reasonable.}
Now that we have characterised the intrinsic shape distribution of each control sample, we use
these measured shapes to estimate the viewing angles of the AGN populations as a whole, and
for different AGN and morphological types separately. 

\section{Inclinations and orientation alignments of the AGN population}
\label{sec:inclinations}

We aim to obtain the inclination angle distribution of the  AGN host
galaxy population
to determine whether there is a preference for face- or edge-on orientations in a given AGN population.
To do this, we first calculate the predicted likelihood distribution of
the projected $b/a$
as a function of the inclination angle ($\theta$), given our model for 
the intrinsic shapes of galaxies of each control
sample (i.e. parameters listed in Table~\ref{ParametersPS08}).
This is shown in Fig.~\ref{anglespop}. Here, the colour bar shows the likelihood
normalized to give an integral of $1$ at each value of $b/a$.
Elliptical galaxies are characterised
by wide distributions in $\theta$ at a given $b/a$
as a result of their intrinsic round shapes, while spiral galaxies have much narrower
distributions, {supported by the value of the probability peak in each
case. Note that there are two local maxima in the top-right and bottom-left of
each panel. These represent a non-null probability of having an intrinsically
round or elongated object which has an edge- or face-on inclination, respectively. This
probability is more significant in the case of elliptical galaxies.}

\begin{figure*}
\begin{center}
\includegraphics[width=0.43\textwidth]{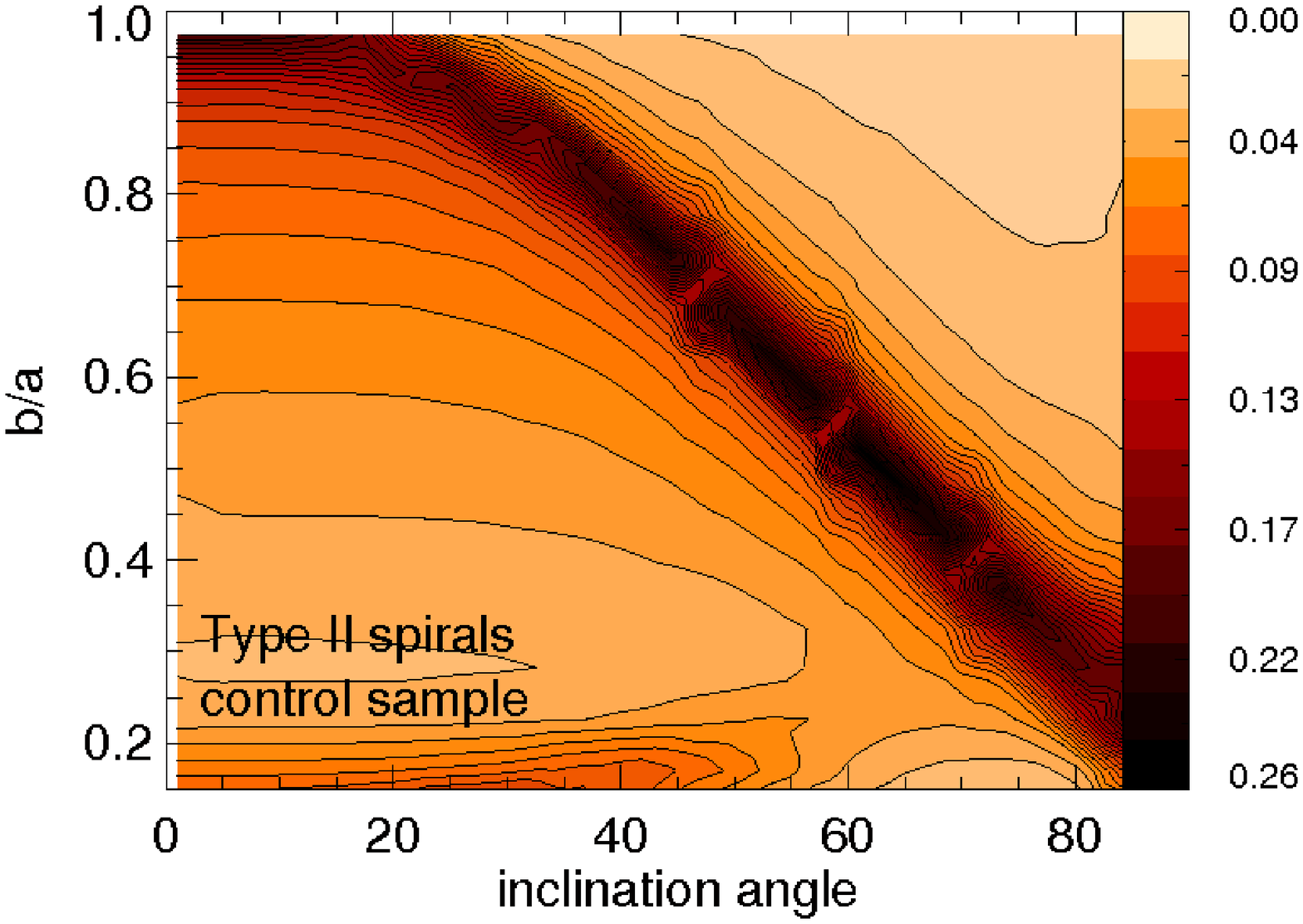}
\includegraphics[width=0.43\textwidth]{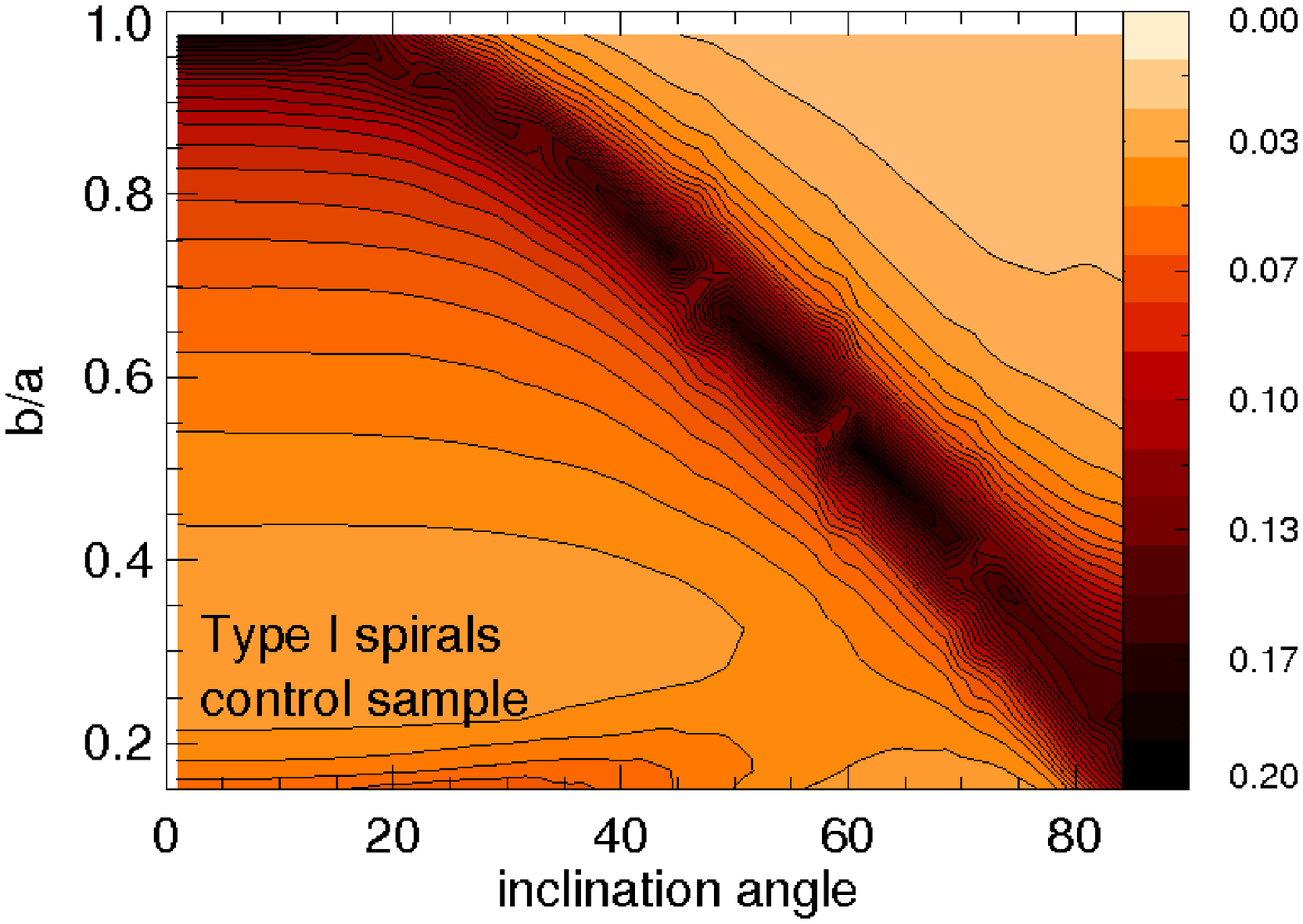}
\includegraphics[width=0.43\textwidth]{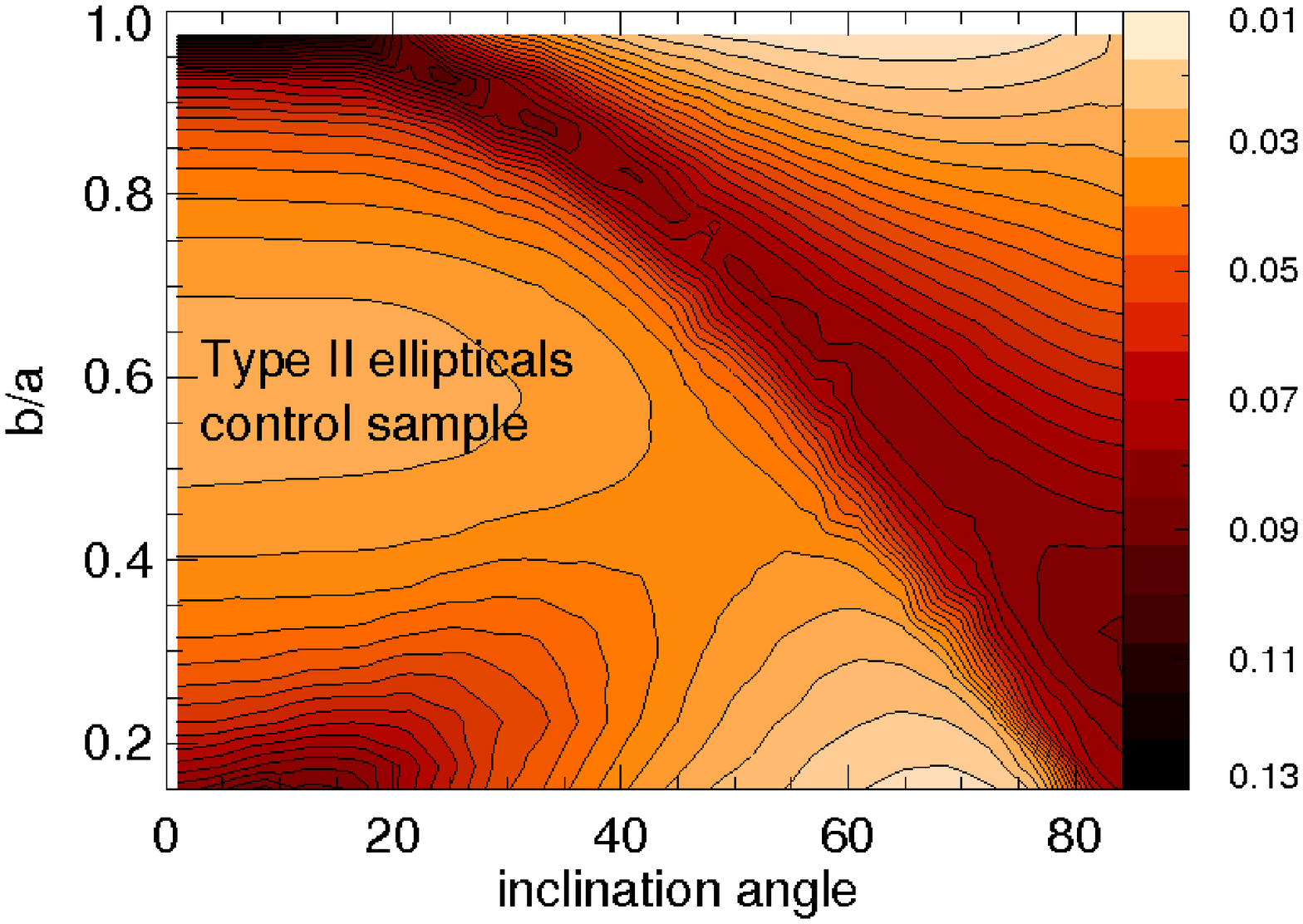}
\includegraphics[width=0.43\textwidth]{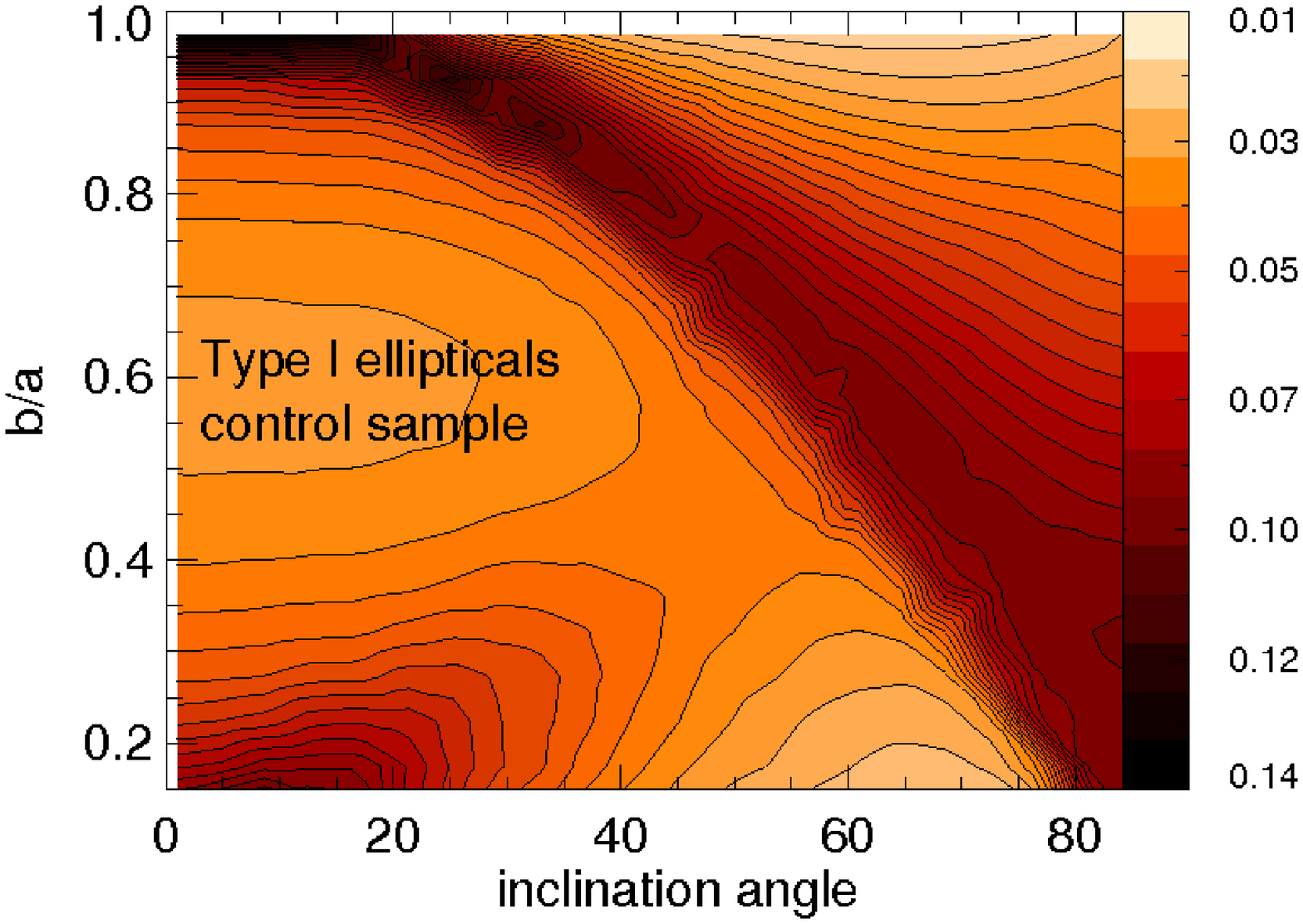}
\caption{\label{anglespop}
Probability density contours in the axis ratio, $b/a$, vs. inclination angle plane for each galaxy
population selected to mimic the luminosity, 
colour and concentration properties of the AGN population indicated in
each panel. These results take into account the effects of dust extinction. 
The scale is shown on the colour bar at the right of each
plot. Probabilities were normalized to give an integral of $1$ in horizontal cuts of $b/a$.}
\end{center}
\end{figure*}

For each control galaxy sample we determine their distribution of inclination
angles using the distribution of
measured $b/a$ values and
the probability function of inclination angles vs. $b/a$
shown in Fig.~\ref{anglespop}.
These probability functions peak at higher
values of inclination angle when $b/a$ decreases, and are narrower at
lower $b/a$ values in all the control samples.
We count the
number of galaxies at each inclination angle, $\theta$, 
weighting by $W(\theta)= V_{\rm max}^{-1} \times W_{\rm Gal}(\theta\,| b/a)$, where
$W_{\rm Gal}(\theta\,| b/a)$ is the 
probability that a given galaxy has an inclination angle of $\theta$ given its $b/a$.
When we do this for the control samples 
we find that they are consistent within the errors with uniform distributions in
$\rm cos(\theta)$, as expected.
To quantify the deviations from uniformity for the AGN sample, we study the ratios between the normalised
distributions of $\rm cos(\theta)$
of the AGN population and the corresponding control sample, $f_{\rm AGN}/f_{\rm
control}$.

Fig.~\ref{Thetas} shows 
the distributions of $f_{\rm AGN}/f_{\rm control}$ 
for spiral (solid) and elliptical (dashed) type II (top) and type I (bottom)
AGN populations. The errorbars
were calculated using the jackknife technique. 
Spiral type I AGNs have a clear tendency to be face-on; the tendency is weaker
but still significant in the elliptical type I population.
Type II ellipticals show the opposite tendency, with a distribution that has a
maximum at $\theta\approx 90^o$, while type II spirals show no clear 
skew in their
inclinations. 
The weak signal in the latter cases 
is consistent with the similarity in the $b/a$ distributions of the AGN and control samples
(i.e. $\Delta_{\rm b/a}\approx 0.05$, Fig.~\ref{AGNIIEll}), 
which can be interpreted in our model as differences in the polar
viewing angle distributions. 
In general, the
results from Fig.~\ref{Thetas} allow us to rule out random orientations (i.e. a
flat distribution in $f_{\rm AGN}/f_{\rm control}$) 
with a confidence of $\delta\,\chi^2\approx230$ in the type I spiral population, 
$\delta\,\chi^2\approx3$ in the type I elliptical population, 
$\delta\,\chi^2\approx15$ in the type II elliptical population, and 
$\delta\,\chi^2\approx6$ in the type II spiral population. 

{There have been several previous attempts to study the $b/a$ distributions of type I and II
AGNs (e.g. \citealt{Keel80}, \citealt{Lawrence82}, \citealt{Kirhakos90}, \citealt{McLeod95}, 
\citealt{Nagar99}, \citealt{Rigby06}), finding that type I AGN usually appear to have axis ratios close to $1$, 
in agreement with our findings. Our approach is the first to include
a consistent model that takes into account the intrinsic shapes of the AGN hosts and their
intrinsic dust extinction.}

\begin{figure}
\begin{center}
\includegraphics[width=0.43\textwidth]{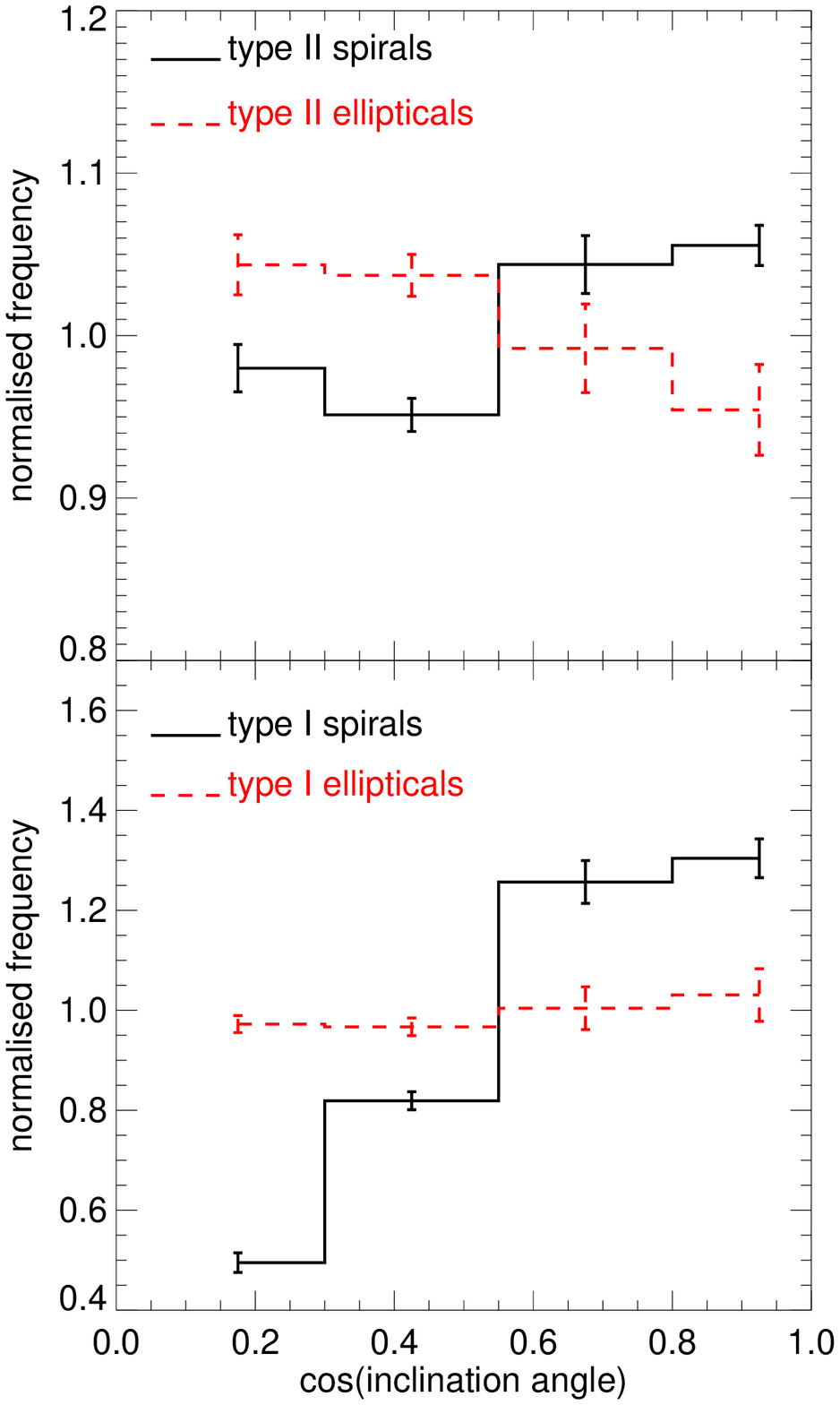}
\caption{\label{Thetas} 
Ratios between normalised frequencies of $\rm cos(\theta)$ 
of the AGN population and their corresponding control samples, $f_{\rm AGN}/f_{\rm control}$,  
for the spiral (solid lines) and
elliptical (dashed lines) type
II (top) and type I (bottom) AGN populations. 
{The distributions are $1/V_{\rm max}$-weighted.}
Errorbars were calculated using the jackknife technique.}
\end{center}
\end{figure}

In the unified model the type of AGN one observes is 
just a matter of the inclination of the obscuring torus 
to the line-of-sight (e.g. \citealt{Antonucci93}). Within this scenario,
{skewness in the distribution of} the inclination angles of host AGN galaxies, like the ones we have found here, 
would reflect a preferential alignment between the large structures of
the galaxy (kpc scales) and their central regions, 
as one would expect in the coherent scenario (see LPC09). 
However, the galactic disc itself may contribute 
to the absorption of the emission from the central engine, and particularly from the
BLR (e.g. \citealt{Maiolino95a}; \citealt{Lutz03}; 
\citealt{Maiolino03}; \citealt{Satyapal08}; \citealt{Goulding09}, 2010), giving
rise to {skewness} in the 
inclination angle distribution of AGN hosts. This is why it is important to explore the nature of the 
observed tendencies towards face- or edge-on of Fig~\ref{Thetas}.

In order to gain insight into the nature of the 
inclination angle distribution of the AGN samples, we study the dependence of the frequency 
shown in Fig~\ref{Thetas} on properties of the
$[\rm OIII]$ emission line. We also calculate the expected 
distributions of $b/a$ for spirals if the galactic disc is
responsible for the absorption of BLR emission in type II AGN.

\subsection{Possible completeness problems in the AGN samples}
\label{ssec:ew}

\begin{figure*}
\begin{center}
\includegraphics[width=0.8\textwidth]{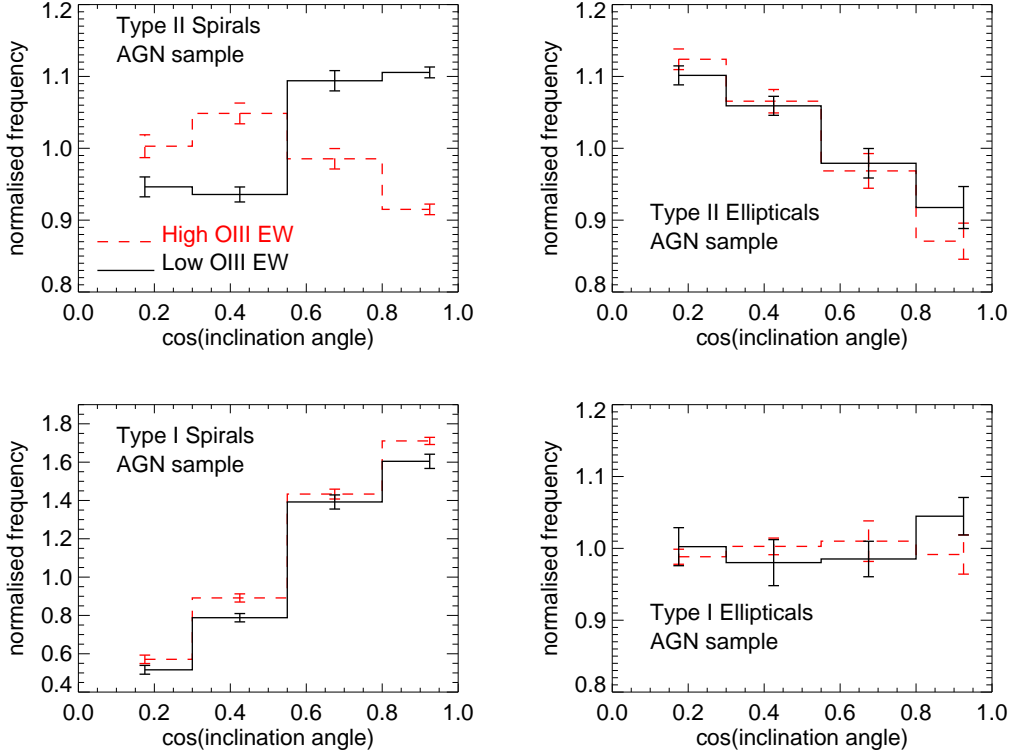}
\caption{\label{ThetasOIII} 
Ratios between normalised frequencies of $\rm cos(\theta)$ 
of the AGN population and the corresponding control samples, 
for the spiral and elliptical 
type II (top) and type I (bottom) AGN populations, {separated 
according to their $\rm [OIII]$ equivalent width (low EW in solid lines and high
EW in dashed lines)}.}
\end{center}
\end{figure*}

Line emission from star formation activity in galaxies could represent an important 
bias in the optical classification of AGN from their emission-line properties. 
Several authors (e.g \citealt{Goulding09},
2010; \citealt{Juneau10}) have shown that a significant population of Seyfert
IIs hosted by edge-on galaxies are misclassified
due to this effect at low and intermediate redshifts. 
We expect this effect to be more important at low emission line 
luminosities since in this regime, the star-formation contribution to the emission-line flux could overwhelm 
the AGN light. Note that faint, low-luminosity emission lines are also difficult
to detect given our high S/N limits. 
In this section we explore the sensitivity of the inclination 
angle distribution to the EW of the lines, with these effects included.

\begin{table}
\begin{center}
\caption{Weighted fraction of edge-on galaxies in the {type I and type II} 
AGN spiral galaxies samples compared
with the corresponding control sample, and in subsamples of high and low $[\rm OIII]$ equivalent
width (see Fig.~\ref{ThetasOIII}). Edge-on galaxies are defined as those having an
inclination angle of $\rm cos(\theta) \le 2 (1-\gamma)$ (i.e. corresponding to a maximum deviation from the line of
sight of $1-\gamma=<C/B>$, see Table~\ref{ParametersPS08}). {Errors indicate the 
standard deviation of the mean.}}\label{FractionEOTIISpi}
\begin{tabular}{c c c}
\\[3pt]
\hline

\small{Sample} & AGN sample & control sample \\
\hline
All Type II AGN spirals& $0.58\pm0.05$  & $0.56\pm0.05$   \\
Low $[\rm OIII]$ EW & $0.59\pm0.07$  & $0.63\pm0.06$ \\
High $[\rm OIII]$ EW & $0.70\pm0.03$ & $0.62\pm0.04$ \\
\hline
All Type I AGN spirals& $0.28\pm0.06$  & $0.44\pm0.05$   \\
Low $[\rm OIII]$ EW & $0.36\pm0.06$  & $0.56\pm0.07$ \\
High $[\rm OIII]$ EW & $0.43\pm0.07$ & $0.62\pm0.05$ \\
\hline
\end{tabular}
\end{center}
\end{table}

We divide the four samples of AGN galaxies in subsamples of low and 
high rest-frame EW of the $\rm [OIII]$
emission line\footnote{{We remind the reader that the $\rm [OIII]$ emission line is generated in the 
narrow line region of AGNs, and therefore always appears narrow independent of the AGN type. 
This enables us to do a fair comparison between type I and II AGN subsamples.}} 
(divided at the sample median $\rm [OIII]$ EW of $\approx 3 \AA$). {For each subsample we
construct its corresponding control sample as in $\S 2$. We find that 
these control samples are characterised by the 
same concentrations and $g-r$ colours as those shown in Fig.~\ref{select1}. However, the typical 
luminosities of the high and low $\rm
[OIII]$ EW samples are brighter and fainter by $\approx 0.3$ magnitudes than the distributions 
of Fig.~\ref{select1}, respectively. This is consistent with the observed 
BH-bulge relations (e.g. \citealt{Ferrarese00}; \citealt{Marconi03}; \citealt{Haring04}), 
where more massive BHs (which drive brighter AGN, in 
this case linked with higher $\rm [OIII]$ EW) are hosted by more massive galaxies. 
Despite this difference in the intrinsic luminosity of the hosts, 
the intrinsic shape parameters derived from the control samples in each
subsample of low and high [OIII] EW agree with each other within the uncertainties, and with the
parameters found for each AGN sample overall (Table~\ref{ParametersPS08}). Interestingly, 
the marginal differences seen between the intrinsic shapes of 
type I and II AGN spirals in Table~\ref{ParametersPS08} 
disappear when comparing the high [OIII] EW subsamples.} 

The resulting $\rm cos(\theta)$ distributions of the subsample of AGN galaxies of high and low $\rm
[OIII]$ EW are shown in Fig.~\ref{ThetasOIII}. 
The low/high $\rm [OIII]$ EW subsamples are in good agreement
with one another within the same AGN sample,
except for type II spiral galaxies, where the high $\rm
[OIII]$ EW subsample shows a strong edge-on
tendency. 

Since the control samples have underlying shapes in agreement with each other, 
the differences seen between the low and high $\rm [OIII]$ EW subsamples of the
same AGN sample in Fig.~\ref{ThetasOIII} are not due to 
intrinsic differences in the hosts.
This suggests that the
low EW [OIII] subsample of type II spirals is subject to larger systematic biases 
than its high EW counterpart, and the unified model does provide a good explanation of
the results for the latter. 

In order to quantify the differences between the low and high $\rm
[OIII]$ EW subsamples of type II AGN spirals, we calculate the $V_{\rm MAX}$-weighted
fractions of edge-on galaxies (those with $\rm cos(\theta)\le 2 (1-\gamma)=2
\,<C/B>$, 
i.e. $\theta \gtrsim 60^o$) in each subsample. 
These fractions are shown in Table~\ref{FractionEOTIISpi}. {For reference,
we also show the edge-on galaxy fraction for type I AGN spirals, and for low
and high [OIII] EW subsamples of the type I spirals.}
This fraction for the high $\rm [OIII]$ EW subsample is $20$\% higher than in
the low $\rm [OIII]$ EW subsample in type II spirals, {and is larger than the
errors of the means}. This suggests we
are missing $\approx 20$\% of edge-on type II spirals with low $\rm [OIII]$ EW.
This number is a lower limit, since it assumes that the high $\rm [OIII]$ EW
subsample is complete and that we are just losing edge-on galaxies. This crude 
estimate is consistent within a factor of $2$ with the estimate by 
\citealt{Juneau10} of optically 
misclassified SDSS DR4 AGN identified by their X-ray emission. 
They found that $40$\% of X-ray-bright edge-on galaxies are 
classified as star forming or composite star forming/AGN galaxies. 

{In the
case of type I AGN spirals, the high [OIII] EW subsample shows a larger fraction
of edge-on galaxies than the low [OIII] EW subsample, possibly indicating
that the bias mentioned above is also affecting the type I AGN sample. However, when compared to
the control sample, the difference is only $8$\%, which is smaller than the
errors. Thus, the two subsamples are consistent within the errors, as expected from the distributions
of Fig.~\ref{FractionEOTIISpi}. We conclude that obscuration from the galactic
disc only has a significant effect on our ability to classify type II spirals.} 

\subsection{Could the broad line absorption come from the galactic disc?}

The obscuration preventing the
direct observation of the accretion disc in type II AGN galaxies
could be produced by a local torus or the galactic disc. 
Even though polarimetry results support the unified AGN model which invokes the local torus,
detailed analysis of nearby galaxies have also revealed 
absorption of broad line regions by the galactic disc in some cases 
(\citealt{Goulding09};
\citealt{Goulding10}). 

\begin{figure}
\begin{center}
\includegraphics[width=0.45\textwidth]{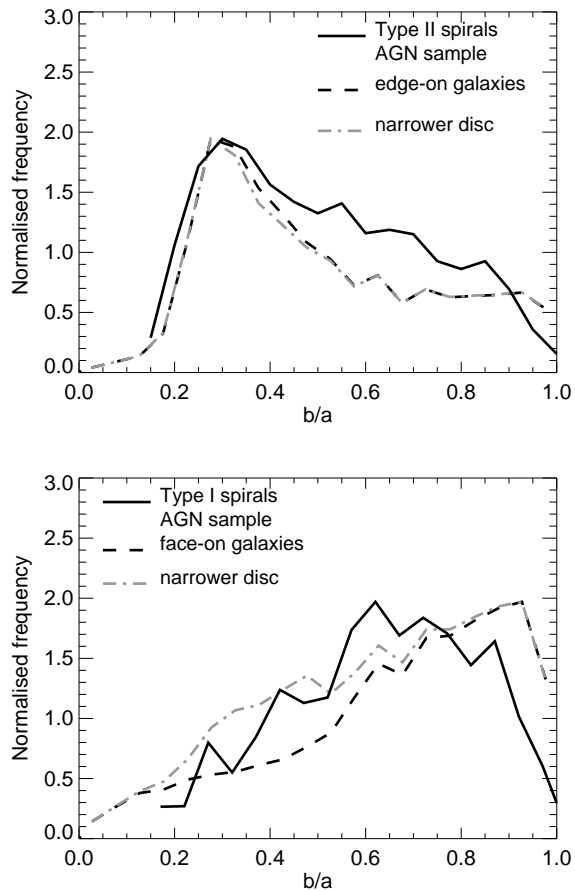}
\caption{\label{CompositebsDis} 
Expected distributions of $b/a$ when the absorption of the broad emission lines comes
from the disc of the galaxies instead of a nuclear torus, normalised
to the number of galaxies in the samples of AGN galaxies (dashed lines).
The distribution of projected axis ratios, $b/a$, for the type II (top
panel) and type I (bottom panel) AGN spirals are shown as solid lines, and the corresponding
control samples as dotted lines. 
{For reference, we also show as
a dot-dashed line the composite $b/a$ distributions for the case where the galactic disc
is half as thick (i.e. smaller $C/B$).}}
\end{center}
\end{figure}

The results of the previous sections do not distinguish between these two possibilities. 
In an attempt to address this question, 
we test whether the observed $b/a$ distributions of type I and II spiral AGN can
be reproduced by absorption by dust in the discs of their host galaxies\footnote{{
It has been pointed out that elliptical galaxies often show circumnuclear discs
which can also obscure the central engine (e.g. \citealt{Kawata07}).
However, our method is not able to probe this possibility.}}. 
To do this we use the heuristic dust model adopted by PS08, which depends on the
parameter $\gamma$, the mean height of the galaxy ellipsoid.
We assume that
the dust lanes produce a covering fraction which 
depends on the luminosity of the galaxy and that varies with the inclination angle $\theta$ of the galaxy as
\begin{equation}
f_{\rm cov}(\theta)= \left \{
\begin{array}{l}
f_0\times10^{0.4 (\gamma-\cos\theta)},  {\rm if} \cos\theta>\gamma \nonumber \\
f_0,  {\rm if} \cos\theta<\gamma 
\end{array} \right.
\label{eq:ext}
\end{equation}  
where $f_0$ is the edge-on covering fraction. Comparing 
with $\S 2$, the $10^{0.4 (\gamma-\cos\theta)}$ factor reflects 
the fact that we are considering extinction in luminosity rather than magnitudes. 
We adopt the extreme case of $f_0=1$. 
We produce distributions of $b/a$ for galaxies with inclination angles in bins
of $\cos(\theta)$ weighted by the corresponding covering fraction $f_{\rm cov}$, add them
together, and compare the result with the observed $b/a$ distribution of type I and II AGN spirals.

{This is shown in Fig.~\ref{CompositebsDis}, where the observed $b/a$
distributions for the AGN samples (see Fig.~\ref{AGNIIEll}) are shown as solid lines, and the composite
distributions for the case of pure disc absorption ($f_0=1$) are shown as dashed lines. 
For reference, we also show
the composite distributions in the case where we consider galactic discs that are narrower, i.e. in $C/B$, 
by a factor of two. The latter choice is motivated by the fact that the PS08 model
attempts to fit the bulge and the galactic disc as a single photometric
structure instead of distinguishing between them, which could spuriously broaden the
galactic disc}. Absorption by the
galactic disc alone {goes in the right direction, but does not fully reproduce} the distributions of either
type I or II AGN even assuming $f_0=1$; the resulting distributions show excesses 
of face-on and edge-on galaxies in the two cases, respectively. By considering narrower galactic discs we are
able to broaden the $b/a$ distributions for type I spiral AGN slightly.
However, the output distribution still has a clear excess of face-on galaxies. 
Moreover, $f_{\rm cov}$ is not strongly dependent on $\theta$ (i.e. a large 
population of type II AGNs could come from 
fairly face-on galaxies), 
meaning that we are forcing the composite $b/a$
distribution to be as broad as possible. Thus a more physical
approach would give even narrower distributions. In addition, the 
pure galactic disc absorption scenario adopted here
produces five times more type
II than type I AGN at a given luminosity, in contradiction with 
the observed ratio of $0.25-0.5$\footnote{Note that
these ratios are $V_{\rm MAX}$-weighted and thus refer to weighted number of detections.} (\citealt{Hao05b}; see also $\S 4$).

Our results indicate that even if part of the broad line absorption takes
place in galactic discs, additional sources of absorption such as a torus, are still needed.
We will bear this in mind when analysing the relative orientations
of torus and galactic disc angular momenta in the following section.

\section{Alignments between the galaxy disc and an obscuring torus}
\label{sec:align}

\begin{figure*}
\begin{center}
\includegraphics[width=1.0\textwidth]{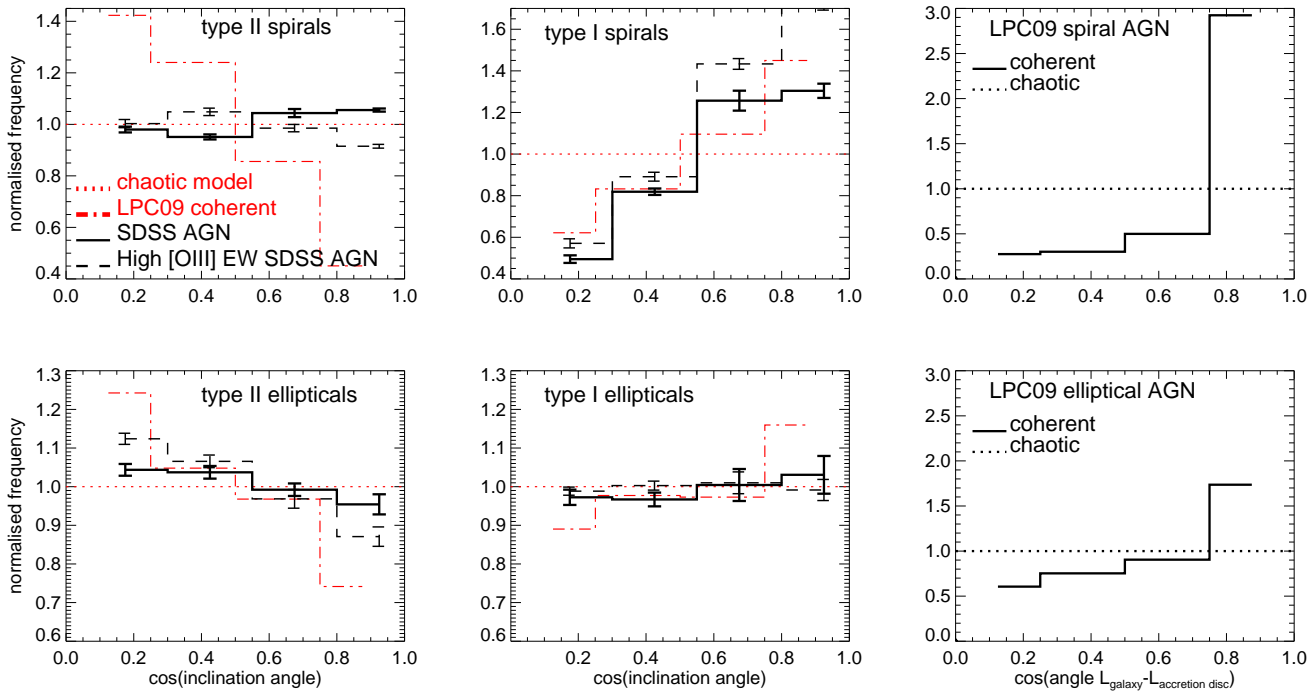}
\caption{\label{CompSDSS} {\it Left and middle panels:} Distribution of the inclination angle of 
AGN host galaxies from Fig.~\ref{Thetas} (in solid lines; AGN types are as labelled) compared to 
the predictions of the semi-analytic model LPC09 in the 
coherent (dot-dashed lines) and chaotic (dotted lines) scenarios. For reference, 
we also show the $\rm cos(\theta)$ distributions for high [OIII] EW subsamples of Fig.~\ref{ThetasOIII} 
as dashed lines.
{\it Right panels:} Predicted distributions of the angle between the angular momentum of the galaxy 
and the accretion disc for the LPC09 in the coherent (solid line) and 
chaotic (dashed line) scenarios for elliptical and spiral galaxies (classified as in \citealt{Lagos08} 
to account for the observed morphology type fractions as a function of 
stellar mass from \citealt{Conselice06}). Errorbars were 
calculated using the jackknife technique.}
\end{center}
\end{figure*}

We use the measured inclination angle distribution (cf. Fig.~\ref{Thetas}) to analyse 
alignments between the galaxy disc and the broad torus in spiral galaxies
{within the framework of the AGN unified model}. 

Within this paradigm, the obscuring torus must be broad in order to explain the
wide inclination angle range over which type II AGN are found, and the ratio between the number
of type II and type I AGN galaxies (e.g. \citealt{Hao05b}; \citealt{Treister08}; \citealt{Reyes08}).  
We quantify the typical torus width by considering that the fraction 
of the solid angle covered by the torus corresponds to the weighted ratio of type II AGN to the
total number of AGN,

\begin{equation}
f^{-1}=1+\frac{\sum W_{\rm AGN TII}}{\sum W_{\rm AGN TI}},
\label{peso}
\end{equation}

\noindent where $W_{\rm AGN}=V_{\rm max}^{-1}$. 
This calculation shows that the torus has a typical azimuthal height
of $\approx 40^o$, which corresponds to a $V_{\rm MAX}$-weighted ratio of type I to II of 3:1.
By measuring the $[\rm OIII]$ luminosity function of type I and II AGNs, Hao et al. (2005b) show
that this ratio depends on the AGN luminosity.  They find ratios close to unity for low luminosities,
$\simeq2:1$ for $10^{5.8}<L(\rm [OIII])<10^{6.5}L_{\odot}$, and $\simeq4:1$ for
$L(\rm [OIII])>10^{6.5}L_{\odot}$, 
broadly consistent with our estimate, considering that the weighted mean luminosity
of our sample is $L([\rm OIII])\approx10^{6.4} L_{\odot}$. 

We use the output galaxies from the semi-analytic model of LPC09 to construct samples 
of modeled type I and type II AGN spirals and ellipticals 
using the typical torus width inferred above. We 
choose a random observer's line of sight and
estimate the projected angle subtended by the angular 
momentum of the accretion disc, $\theta_{\rm acc}$. We assume  
perfect alignment between the obscuring torus 
and the accretion disc\footnote{This is expected at least to
a high degree in models where the absorption of BLR occurs in radiation-driven winds from 
the accretion disk (see for instance \citealt{Murray95}).}. We run a 
Monte-Carlo simulation and extract torus heights, $h_{\rm t}$, so that the mean corresponds to 
$\tan(40^o)\approx 0.83$. We consider those galaxies with $\theta_{\rm
acc}+h_{\rm t}>\pi/2$ as type II AGN, and the rest as type I AGN. 
Given each galaxy angular momentum direction, we estimate the observed projected
inclination angles.

The left and middle panels of Fig.~\ref{CompSDSS} show the resulting normalised 
frequency of the inclination angle of AGN hosts for the LPC09 semi-analytic model in the 
coherent and chaotic scenarios, and compare this
with the observed distributions from Fig.~\ref{Thetas}. For reference, we also show the inclination angle distributions 
for the high [OIII] EW subsamples from Fig.~\ref{ThetasOIII} since they are more 
likely to be complete samples (see $\S 3.1$). The right panel of Fig.~\ref{CompSDSS} 
shows the theoretical expectations for the angle between the angular momentum of the galaxy and 
the accretion disc in the coherent and chaotic scenarios.
In the chaotic model, the two angular momenta are unrelated and the distribution is flat. 
In the coherent model they are highly aligned due to the assumption that the accretion 
flow to the BH is perfectly aligned with the large-scale source of the gas. 
The coherent scenario reproduces the observed distributions 
of type I AGN quite closely, but it fails for type II AGN (particularly for type
II AGN spirals). The discrepancies decrease slightly
when comparing to the high [OIII] EW subsamples.

\begin{figure}
\begin{center}
\includegraphics[width=0.4\textwidth]{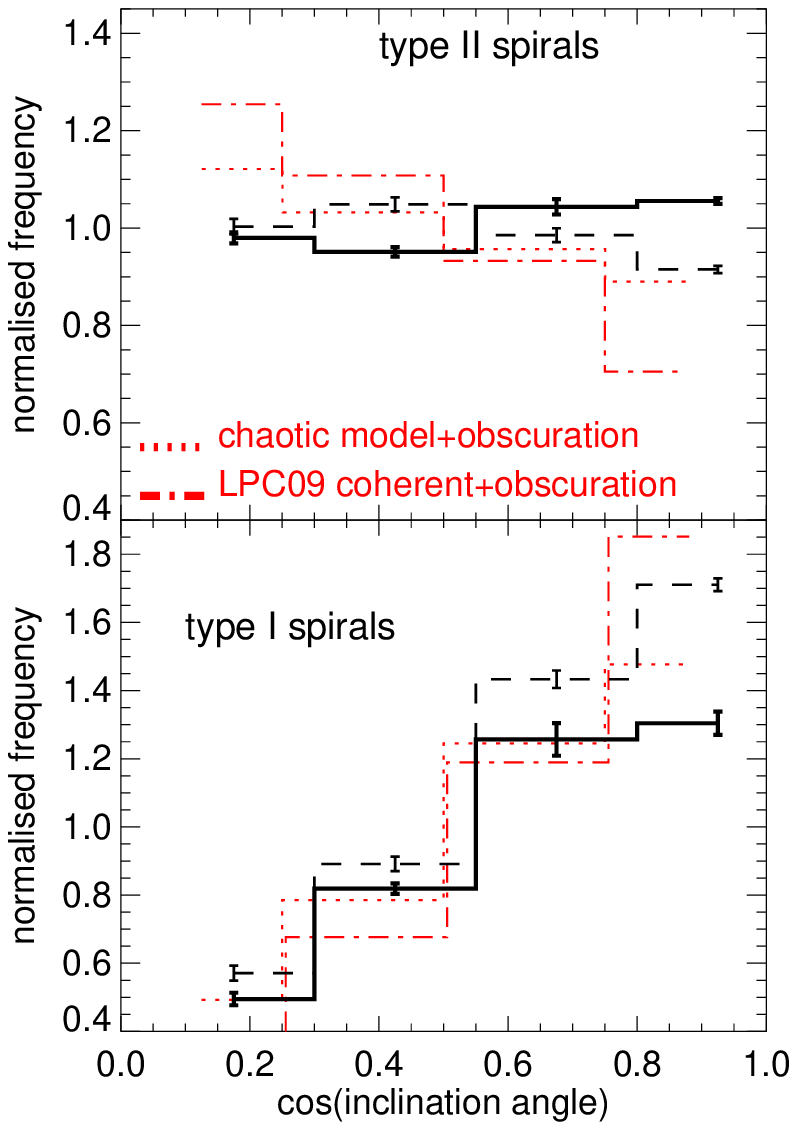}
\caption{\label{CompSDSSobs} Distribution of the inclination angle of 
AGN host spiral galaxies from Fig.~\ref{Thetas} (in solid lines; AGN types are
as labelled) and for the high [OIII] EW subsamples of Fig.~\ref{ThetasOIII}
(dashed lines), compared to 
the predictions of the semi-analytic model LPC09 in the 
coherent (dot-dashed lines) and chaotic (dotted lines) scenarios, when the 
effect of galactic disc obscuration, by using the heuristic 
model of $\S 3.2$, is included.} 
\end{center}
\end{figure}     

{As was mentioned above, obscuration by the galactic disc may also affect the
determination of the AGN type in spiral
galaxies (see $\S 3$). We consider this possibility by also including the heuristic
model for the obscuration of the AGN emission by the galactic disc of $\S 3.2$ 
in the predictions of the LCP09 model in the coherent and chaotic scenarios. In
this case, highly inclined spiral galaxies have higher probability of being seen
as type II AGN. A fraction $f_{\rm cov}(\theta)$ of spiral galaxies at inclination $\theta$ will
be classified as type II AGN. Fig.~\ref{CompSDSSobs} shows the distribution of $\rm cos(\theta)$ for
spiral galaxies from Fig.~\ref{CompSDSS}, together with the chaotic model
and the coherent model when the obscuration from the disc is included. The tendency of
alignment observed
in the LPC09 model in the coherent scenario becomes weaker for the type II
spirals, and stronger in 
type I spirals. A tendency towards edge- and
face-on orientations appears for type II and I AGN, respectively, in the chaotic
model. This tendency
is very similar to that observed: a very weak edge-on tendency for type II
spirals, and a much stronger tendency towards face-on orientations for type I
spirals. This suggests
that the observed stronger signal for a preferred
inclination of type I
compared to type II hosts may be driven by obscuration from the galactic disc.
However, the model of galactic disc 
obscuration we have used is maximal in the sense that it assumes that edge-on AGN galaxies are observed as type II AGN. 
More importantly, 
the adopted disc obscuration cannot reproduce the distribution of 
projected shapes of AGN hosts (see Fig.~\ref{CompositebsDis}) or explain the
tendencies observed in the elliptical population. Thus, disc obscuration cannot
fully explain the type I and II dichotomy reported here.

On the other hand, the coherent model predicts a 
higher {skewing} towards face- or edge-on orientations of AGN hosts than we
see in the observations, 
indicating that the degree of alignment between the galaxy and the torus (or
accretion disc), in the 
observational sample is much weaker than in the idealized coherent scenario.

We summarize the comparison between the observed $\rm cos(\theta)$ distribution
of each AGN sample and the predictions of the LPC09 model with and without
disc obscuration in Table~\ref{CompAllMods}. As can be seen from the $\delta \chi^2$ between observations and models,
neither, the
chaotic nor the coherent model, is able to closely reproduce the
observed distributions. The only exception is the high [OIII] EW subsample of
type I AGN ellipticals, which is consistent with random orientations. 
This suggests that in general, the coherent and the chaotic scenarios represent
idealized conditions;
i.e. some loss of the angular momentum direction is expected during the infall,
but this cannot be complete.} 

\begin{table*}
\begin{center}
\caption{{$\delta \chi^{2}=\chi^{2}/d.o.f.$ from the comparison between the inferred distribution of
inclination angles of AGN hosts in the SDSS and the predictions of the LPC09
model in three variants: (i) the chaotic scenario, (ii) the coherent
scenario (see Fig.~\ref{CompSDSS}), and
(iii) the coherent and chaotic scenarios considering obscuration from the 
galactic disc (only applied to spiral galaxies; see Fig~\ref{CompSDSSobs}).}}\label{CompAllMods}
\begin{tabular}{c c c c c}
\\[3pt]
\hline
\small{Sample} & $\delta \chi^{2}$ chaotic & $\delta \chi^{2}$ coherent &
$\delta \chi^{2}$ coherent+obscuration & $\delta \chi^{2}$ chaotic+obscuration\\
\hline
Type I AGN spirals & 229 & 20 & 137 & 7 \\
Type II AGN spirals &  6 & 2007 & 700 & 250\\
Type I AGN ellipticals &  3 &8  & - & -\\
Type II AGN ellipticals &  15 & 86 & - & -\\
\hline
Type I AGN spirals high [OIII] EW& 535 & 94 & 100 & 61 \\
Type II AGN spirals high [OIII] EW& 37 & 1236 & 270 & 18\\
Type I AGN ellipticals high [OIII] EW& 0.5 & 33 & - & - \\
Type II AGN ellipticals high [OIII] EW&  30 & 23 & - & -\\
\hline
\end{tabular}
\end{center}
\end{table*}

Our results indicate that, within the AGN unified model framework, there is some
weak coherence 
between the torus and the host galaxy angular momenta, and that the purely chaotic scenario is ruled out.

\section{Discussion}

The results shown in the last two sections point to a scenario in
which the AGN components, i.e. torus and accretion disc, are aligned to a significant
degree with the angular momentum of the host galaxy. 

Theoretical suggestions by LPC09 indicate that, under this condition, high BH spin values
would be commonplace (cf. Section~$1$). This preference for
alignments could be explained by short timescales characterising the 
gas infall to the central parts of the galaxy.
\citet{Bogdanovic07} show that this is likely to happen in gas-rich
mergers (see also \citealt{Barausse09}), 
where the BHs taking part in the galaxy merger will acquire
$1$\%-$10$\% of their mass in a time which is short compared to the time needed for
the BHs to spiral in towards the center ($\le 5\times 10^7$~yr,
\citealt{Escala04}, 2005), or the 
time for a starburst to deplete the supply of gas ($\approx 10^8$~yr,
\citealt{Larson87}). A potential problem with the 
preservation of the angular momentum
direction in inflowing gas assumed in the coherent model comes from the
difficulty in transporting gas through the corotation resonance radii of galaxies
(\citealt{Zhang07}). However, \citet{Haan09} found evidence for gas inflow beyond
this radius in seven galaxies in the sample studied by \citet{Garcia-Burillo03},
which continues to their resolution limit, $100$pc away from the nuclei. 

Direct comparison with detailed simulations of
the feeding of BHs (e.g. \citealt{Hobbs10}; \citealt{Power10};
\citealt{Hopkins10}) represents an
attractive tool to test accretion scenarios and compare their 
expectations with the orientation results 
presented in this work. Nonetheless, BH feeding is indeed a challenging issue; 
understanding it depends strongly on modeling the local 
environmental effects such as star formation activity,
supernovae explosions, local winds, and so on (e.g. \citealt{Ciotti07}; \citealt{King08};
\citealt{Nayakshin10}).

\section{Summary and conclusions}\label{conclusion}

We study the intrinsic shapes of SDSS DR7 AGN host galaxies
and their inclination angle distributions,
under the assumption that AGN host
galaxies have the
same underlying shapes as normal galaxies matched in
$g-r$ colour, $r-$band luminosity, and concentration, $r_{50}/r_{90}$.  We examine elliptical and
spiral type I and II AGN galaxies separately, and characterise the
$3$-D shapes of
these samples using the model of \citet{Padilla08}.  With these models
we infer
the distributions of inclination angles of AGN hosts. The main
results are as follows.

(i) {The structural parameters of the AGN control samples (Table~$1$)
are consistent with the full SDSS DR6 spiral and elliptical population reported
by \citet{Padilla08}, and correspond to oblate
spheroids with typical $\gamma=<1-C/B>$ of $\approx 0.40 \pm 0.24$ 
for elliptical galaxies, and $\approx 0.77 \pm 0.08$ for spiral galaxies.}
The unified AGN model states that type I and II AGN are similar objects
seen with different orientations. Consistent with this, we find that 
the intrinsic shapes of type I
and II AGN hosts of a given morphological type are comparable.

(ii) Using the intrinsic shape parameters of control samples, we find that
type I AGN galaxies have a strong tendency to be face-on, while type II AGN
galaxies have only a slight tendency to
be edge-on. In particular, type II spiral galaxies show a more uniform inclination angle
distribution consistent with random orientations; only galaxies in the subsample of high [OIII] EW show
clear edge-on orientations, indicating a possible selection effect acting on the low [OIII] EW spiral AGN.  
In the case of type I galaxies we are 
able to rule out random orientations with a confidence of 
$\delta\,\chi^2\approx3$ and $\delta\,\chi^2\approx230$ in the elliptical and spiral 
populations, respectively. In the case of the type II elliptical and spiral 
galaxies, we rule out random orientations with a confidence of
$\delta\,\chi^2\approx15$ and $\delta\,\chi^2\approx6$, respectively. %

(iii) We use the estimated three-dimensional shapes of spiral galaxies to test whether
dust extinction in the galactic discs could be responsible for the absorption of
broad AGN lines in type II AGN galaxies.  We
find that the resulting predicted $b/a$ distributions are not compatible with those observed for either
type I or II AGNs. This indicates that galactic discs cannot be the only source of broad line
absorption, and supports the existence of broad tori surrounding the active nuclei.

(iv) Using the weighted frequency of type I to type II AGN, we found
that the torus producing the absorption of broad lines has a typical
azimuthal height of $\approx 40$ degrees.  

(v) We compare the observed inclination angle distributions with the theoretical predictions of
the LPC09 semi-analytic model for the coherent scenario under the AGN unified
model (e.g. \citealt{Fanaroff74};
\citealt{barthel89}; \citealt{madau94}; \citealt{gunn99}). We find that the model
overpredicts the observed {skewing} towards face- or edge-on
orientations of AGN hosts, but sucessfully reproduces
the differences in the {skewing} between ellipticals and spirals. On
the other hand, the chaotic scenario predicts tendencies towards face- and
edge-on orientations of the type I and II AGN spirals, respectively, only 
if absorption from the galactic disc is assumed. {However, 
in the case of elliptical galaxies 
the chaotic scenario fails unless large amounts of dust aligned with 
the major axis of the galaxy are assumed.}
Our results suggest that some, but not all, of
the direction of the angular momentum of the material
falling to the nucleus is coherent.

These results 
put important constraints on the
physical processes involved in the gas inflow from the outer parts of the galaxy
to the central engine. 
If the coherence of the accretion flow is frequent and large enough, 
high spin values ($\hat{\rm a} \sim 1$) would be commonplace at the massive end of the BH
population (\citealt{Lagos09}), regardless 
of other physical processes as warps \citep{King05} and fragmentation inside the
accretion disc \citep{King08}. {However, a scenario where just a small fraction of the gas flow 
keeps its original angular momentum direction has not been studied theoretically yet, and its consequences 
on BH growth and its spin 
remain to be explored.} 

Kinematic studies of individual 
AGN galaxies, as in \citet{Dumas07}, can help to further explore 
the alignments between gas infall and the whole galaxy.  
It should also be borne in mind that
our approach of fitting the intrinsic shapes of galaxies
from the photometry in SDSS has its limitations; a possible improvement could come from using
the Galaxy Zoo Project (\citealt{Lintott08}) morphologies, which can help to obtain more precise
intrinsic shapes and, therefore, better and more
reliable distributions of inclination angles for AGNs of different types. {A
different and promising approach based on the study of the orientations of host
galaxies of radio sources, selected as to indicate relativistic jets
pointing close to the line of sight, would give
insight on the alignment between the galaxy and
the BH spin (Lagos et al. in prep.).}

Looking further to the future, the LISA survey (\citealt{Johann08}) along with new 
high-resolution X-ray spectroscopy, will provide unequivocal information
on the typical spins of the BH population {and the orientations of jets. This will be
complemented with detailed kinematic studies of the gas in large samples of galaxies (revealing the gas inflows) 
that the Square Kilometer Array (SKA; \citealt{Schilizzi08}) will achieve, helping
to solve these still open questions.} 

\section*{Acknowledgements}

We kindly thank Carlton Baugh, Dave Alexander, St\'ephanie Juneau, 
Philip Best, Chris Power, Peter Creasey and Yetli Rosas-Guevara for useful
comments and discussions. 
We acknowledge the annonymous Referee for
helpful remarks that allowed to
improve this work. CL gratefully acknowledges an STFC 
Gemini studentship. CL and NP were supported by FONDAP 'Centro de Astrof\'isica', BASAL-CATA, and 
FONDECYT No.~1071006. MAS was supported by NSF grant AST-0707266. 
For this project, 500 hours of cputime from the geryon cluster at AIUC were used.

Funding for the SDSS and SDSS-II has been provided by the Alfred P. Sloan Foundation, the Participating Institutions, the National
Science Foundation, the US Department of Energy, the National
Aeronautics and Space Administration, the Japanese Monbuka-gakusho, the Max Planck Society and the Higher Education Funding
Council for England. The SDSS web site is http://www.sdss.org/.
The SDSS is managed by the Astrophysical Research Consortium for the Participating Institutions. The Participating Institutions are the American Museum of Natural History, Astrophysical
Institute Potsdam, University of Basel, University of Cambridge,
Case Western Reserve University, University of Chicago, Drexel
University, Fermilab, the Institute for Advanced Study, the Japan
Participation Group, Johns Hopkins University, the Joint Institute
for Nuclear Astrophysics, the Kavli Institute for Particle Astrophysics and Cosmology, the Korean Scientist Group, the Chinese
Academy of Sciences (LAMOST), Los Alamos National Laboratory, the
Max-Planck-Institute for Astronomy (MPIA), the Max-Planck-Institute for Astrophysics (MPA), New Mexico State Uni-
versity, Ohio State University, University of Pittsburgh, University
of Portsmouth, Princeton University, the United States Naval Observatory and the University of Washington. 
We have also used the FIRST
radio data obtained with the NRAO VLA. The National
Radio Astronomy Observatory is a facility of the National
Science Foundation operated under cooperative agreement
by Associated Universities, Inc. This research has made use
of the NASA/IPAC Extragalactic Database (NED) which
is operated by the Jet Propulsion Laboratory, California
Institute of Technology, under contract with the National
Aeronautics and Space Administration.

\bibliographystyle{mn2e}
\bibliography{Lagos}

\label{lastpage}

\end{document}